\documentclass[12pt,a4paper]{article}
\usepackage{geometry}
\usepackage[round]{natbib}
\usepackage{mathtools}
\usepackage{amsfonts}
\usepackage{amssymb}
\usepackage{graphicx}
\usepackage{stmaryrd}
\usepackage[title]{appendix}
\usepackage[labelfont=bf]{caption}
\usepackage[skip=-0.25\textwidth]{subcaption}
\usepackage{enumitem}%to modify item bullets
\usepackage{titlesec}
\titlelabel{\thetitle.\quad} % to put a "." after section number
\usepackage{xpatch}
\xpretocmd{\eqref}{equation\,}{}{}
\usepackage{authblk}
\title{\textbf{Sedimentation of a surfactant-laden drop under the influence of an electric field}}
\author[1]{Antarip Poddar}
\author[1,2]{Shubhadeep Mandal}
\author[1]{Aditya Bandopadhyay\thanks{Email: aditya@mech.iitkgp.ernet.in}}
\author[1]{Suman Chakraborty\thanks{Email: suman@mech.iitkgp.ernet.in}}

\affil[1]{Department of Mechanical Engineering, Indian Institute of Kharagpur, Kharagpur, West Bengal - 721302, India}
\affil[2]{Max Planck Institute for Dynamics and Self-Organization, Am Fassberg 17, D-37077 G\"{o}ttingen, Germany}
\date{}                     
\begin{document}

\maketitle

\begin{abstract}
  The sedimentation of a surfactant-laden deformable viscous drop acted upon by an electric field is considered theoretically.
  The convection of surfactants in conjunction with the the combined effect electrohydrodynamic flow and sedimentation leads to a
  locally varying surface tension, which subsequently alters the drop dynamics via the interplay of Marangoni, Maxwell and hydrodynamic stresses.
  Assuming small capillary number and small electric Reynolds number, we employ a regular perturbation technique to solve the  coupled system of governing equations.
  It is shown that when a leaky dielectric  drop is sedimenting in another leaky dielectric fluid, Marangoni stress can oppose the  electrohydrodynamic motion severely, thereby causing  corresponding changes in internal flow pattern.
  Such effects further result in retardation  of  drop settling velocity,
  which would have otherwise increased due to the influence of charge convection. For highly mobile surfactants (high P\'eclet number limit),
  the drop surface becomes immobilized and the charge convection
  effect gets completely eliminated. For non-spherical drop shapes, the effect of Marangoni stress is overcome by the `tip stretching' effect on the flow field. As a result, the drop deformation gets intensified with increment in sensitivity of surface tension to the local surfactant concentration. Consequently, for oblate type of deformation the elevated drag
  force causes further reduction in velocity.  Owing to similar reasons, prolate drops experience
  lesser drag and settles faster than the surfactant-free case. In addition to this, with increased
  sensitivity of interfacial tension on the surfactant concentration, the asymmetric deformation
  about the equator gets suppressed. These findings may turn out to be of fundamental significance towards designing electrohydrodynamically actuated droplet based microfluidic systems that are intrinsically tunable by varying the surfactant concentration.  
	
\end{abstract}
\section{Introduction}

It has long been identified that nonlinear effects such as shape distortion, inertia and viscoelasticity lead to enhanced drop manipulation, which have found applications in diverse domains such as drug delivery \citep{Pethig2013}, cell manipulation and separation \citep{Shields2015,Mazutis2013}, protein crystallization \citep{Zheng2004}, ink-jet printing \citep{Basaran2002}, electrohydrodynamic mixing of reactants \citep{Yeo2006} etc. Moreover, control of such effects in a  precise manner is possible in the presence of external fields which includes optical and acoustic waves, temperature gradient  
\citep{Young1959,Borhan1992,Kim1988,Das2017,Sharanya2015}
 and electric field \citep{Taylor1966,Ajayi1978,Feng1999,Lac2007,Xu2006,Esmaeeli2011,Thaokar2012,Mandal2016}.  Application of these external fields can redistribute the interfacial forces and subsequent alteration in droplet motion can be achieved.
 
  An important aspect of the electrical field transport of droplets is the changed patterns of internal flow structures through the appearance of micro-vortices \citep{Tsukada1993,Xu2006}. Thus, eletrohydrodynamics (EHD) has emerged as a widely adopted means for mixing of chemical compounds in droplet based microfluidics \citep{Hoburg1977,Zeng2004,Yeo2006,Tsouris2003}.
 Most of the earlier works related to drop movement under electrohydrodynamic effects considered fluid pairs with nearly equal densities or small enough drop sizes resembling the neutrally buoyant condition.  \citet{Spertell1974} were the first to theoretically predict the coupling behaviour of gravitational settling of drops under the action of an electric field. Comprehensive experiments were performed much later by \citet{Xu2006} and \citet{Ervik2018}. In these cases, the electrohydrodynamic forces affect the drop velocity by altering the surface forces while the gravity acts as a body force on the drop. It has been shown that the resulting drop velocity and shape deformation are due to the coupled effects of these two driving mechanisms \citep{Xu2006, Mandal2016, Mandal2017}. Depending on the electrical properties of the drop and the surrounding fluid, the drop settling velocity may either enhance  or get reduced. In their theoretical calculations, \citet{Xu2006} considered only a marginal shape deformation and charge convection, parametrizing them with a small capillary number ($ Ca $) and a samall electric Reynolds number ($ Re_{\!_E} $), respectively.  A different limit of the same problem was solved after a decade by \citet{Yariv2016} in which they considered the case when the role of the settling velocity is much stronger than the electrohydrodynamic velocity scale.

 Among one of the most rigorously studied interfacial effects in relation to bubbles and drops is the one induced by  a nonuniform distribution of surface-active species on the interface. The surface-active agents (or surfactants) can remain on the interface as contaminants or deliberately used  sometimes \citep{Baret2011,Anna2016}. They  are generally categorized as amphiphilic molecules with ability to make the interface more rigid with decreased mobility of fluid in those regions. With a subsequent movement of the droplet, the surfactant molecules on the interface try to redistribute themselves with a resulting  gradient in surface tension. This initiates a flow opposite to the main flow direction, widely termed as the Marangoni flow \citep{Subramanian2001,Leal2007}. Following the classical work of 
 \citet{Levich1962} on the motion of a bubble under Marangoni effects, a considerable amount of effort has been directed to understand how the migration and deformation characteristics  of drops and bubbles under the action of a background flow can be affected by such effects \citep{Flumerfelt1980,DeBruijn1993,Stone1990,Milliken1994,Sadhal1983,Johnson1983,Hanna2010,Pak2014,Li1997,Stebe1991,Pawar1996}.
  Based on the relative importance of the driving mechanisms of surfactant transport, researchers have employed either or both the non-diffusing and small surface convection limits while obtaining solutions for the equation governing the surfactant distribution under an applied flow field \citep{Hanna2010,Pak2014,Das2017a,Li1997}.

For a contaminant free drop, it has been known that the transport characteristics and deformation behaviour  are severely affected in the presence of electric field. Depending on specific combination of electrical conductivity ratio ($ R $) and permittivity  ratio ($ S $), the drop can deform to a oblate  or prolate spheroid shape (elongation of droplet perpendicular or parallel to the electric field, respectively) \citep{Taylor1966,Ajayi1978,Tsukada1993,Xu2006,Lanauze2015,Mandal2016,Sengupta2017}. 
It is only recently that the role of interfacial Marangoni effects on the electrohydrodynamic effects has been looked into. The effect of surfactants  on the electro-deformation of neutrally buoyant drops was investigated by experimental techniques \citep{Ha1998} and the corresponding theoretical models were also presented \citep{Ha1995,Teigen2010,Nganguia2013}. \citet{Zhang2015}  studied the  role of nonionic surfactant on the deformation and breakup characteristic of a water droplet in a surrounding oil medium. Later, \citet{Ervik2018}  showed how the insoluble surface active agents can affect the oscillating nature of droplets under AC electric field. In both the studies, the fluid pairs considered had large differences in electrical conductivities. Under such a condition, the effect of charge convection is not present due to non-existence of a tangential component of electrical field \citep{Mandal2016b}.

So far, the studies have concentrated on the sedimentation dynamics of drops in the presence of electric field. The effect of additional Marangoni stresses, complicated by the coupled, non-trivial, non-linear  interplay among the surface flow, surfactant distribution and Maxwell stress, has not been addressed in the literature.
In the present study, we consider steady  motion of a surfactant coated drop which falls under gravity and simultaneously acted upon by a vertical DC electric field. To capture the electric field effects on the fluid flow, we adopt the Taylor-Melcher leaky dielectric framework  \citep{Taylor1966,Melcher1969}. It is assumed that both the fluids are electrically conducting while the charge relaxation time scale is small relative to convective time scale. Further, to capture the effects of shape distortion and charge convection at the interface, we follow a double asymptotic analysis with capillary number ($ Ca $) and electric Reynolds number ($ Re_{\!_E} $) as small perturbation parameters, respectively. Also, the surfactant transport equation is considered in the two limiting conditions with low and high surface P\'eclet numbers, respectively. It is worth to mention here that in contrast to the two recent experimental works \citet{Zhang2015, Ervik2018}, we focus on a case of a leaky dielectric drop in another leaky dielectric medium and also highlight the effects of surfactants on the settling velocity. Subsequently, we compare our  results with previous theoretical and numerical results. 
 
\section{Mathematical formulation}
\label{sec:mathform}
\subsection{Governing equations and boundary conditions}
\label{sec:math_gov_eq}
\begin{figure}[!htbp]	
		\centering
	\includegraphics[width=0.5\textwidth]{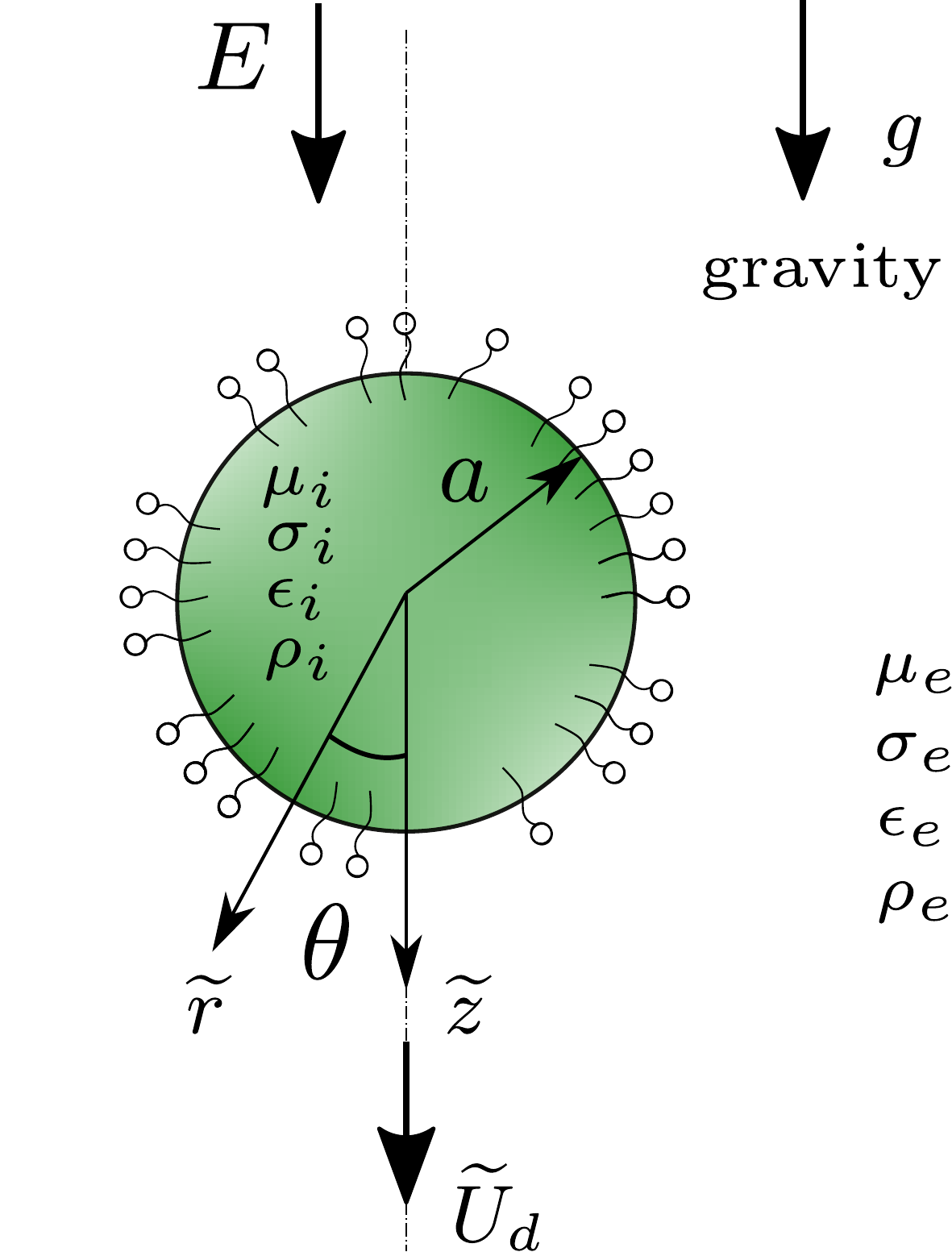}
		\vspace{-5pt} 
	\caption{Schematic representation of the physical setup of a Newtonian drop of undeformed radius $a$ settling with a uniform velocity $\widetilde{U}_d$ in a surrounding unbounded Newtonian fluid medium under the combined action of gravity and a uniform DC electric field $E\,\mathbf{e}_z$. An axisymmetric co-ordinate system ($\widetilde r,\theta$), which is fixed to the centroid of the drop, is used for the present study. The direction of drop motion i.e.\ the $\widetilde z$ - axis is chosen to be aligned with the direction of gravitational field.}
	\label{fig:schematic}
\end{figure}

We consider a physical scenario where a viscous drop of radius $ a $ is sedimenting in  a viscous medium  with an uniform velocity $\mathbf{\widetilde{U}}_{d} \, \mathbf{e}_z$ and acted upon by a DC electric field as shown in figure\,\,\ref{fig:schematic}.  The drop surface is  considered to be covered with nonionic surfactant molecules. Physical properties such as viscosity ($ \mu$), density ($ \rho $), electrical conductivity ($ \sigma $) and dielectric permittivity ($ \epsilon $)  are designated with a subscript `\textit{i}' inside the drop, while a subscript `\textit{e}' is used for the corresponding properties in the outside medium. In the absence of any flow field the drop interface is uniformly coated with surface concentration $ \widetilde{\Gamma}_{eq} $. In the presence of a surface flow such uniformity is destroyed and an axisymmetric concentration distribution $ \widetilde{\Gamma}(\theta) $ is achieved leading to a nonuniform distribution of surface tension $ \widetilde{\gamma}({\theta}) $. The deformed shape is represented as ($ \widetilde{r}_{\!_S}(\theta)$).

\subsubsection{Surfactant transport}
\label{sss:surf trans}
For an isothermal process the interfacial tension depends on the concentration of surfactant molecules on the drop surface. We consider a linear constitutive behaviour between surface tension and the surfactant concentration \citep{Leal2007,Li1997,Stone1990,Nganguia2013,Das2017a}. The dimensional form of the same is given as 
\begin{equation}
\label{eq:linear}
\widetilde{\gamma}(\widetilde{\Gamma})=\widetilde{\gamma}_c-R_gT\,\widetilde{\Gamma},
\end{equation}\\
where $\widetilde{\gamma}_c$, $R_g$ and $T$ are the interfacial tension of the surfactant-free or clean drop, the ideal gas constant and the absolute temperature, respectively.  At equilibrium, for a uniformly coated drop, the interfacial tension decreases can be expressed as $\widetilde{\gamma}_{eq}=\widetilde{\gamma}_c-R_gT\,\widetilde{\Gamma}_{eq}$, where $ \widetilde{\Gamma}_{eq} $ denotes the concentration of a surfactant when the drop surface is uniformly coated. Under such a condition, the \eqref{eq:linear} becomes
\begin{equation}
\label{eq:mod_linear}
\widetilde{\gamma}(\widetilde{\Gamma})=\widetilde{\gamma}_{eq}+R_gT(\widetilde{\Gamma}_{eq}-\widetilde{\Gamma}).
\end{equation}
Assuming a bulk-insoluble surfactant \citep{Leal2007,Stone1990,Li1997} the surfactant distribution at the drop surface is governed by 
\begin{equation}
\widetilde{\nabla}_{\!_S}\cdot(\widetilde{\mathbf{u}}_{\!_S}\widetilde{\Gamma})=D_{\!_S}\widetilde{\nabla}^2_{\!_S}\widetilde{\Gamma}, \qquad \text{at} \,\,\, \tilde{r}=\tilde{r}_{\!_S}(\theta)
\label{eq:dim_surf_transport}
\end{equation}
where $\widetilde{\mathbf{u}}_{\!_{\,S}}=\widetilde{\mathbf{u}}_i|_{\widetilde{r}=\widetilde{r}_{\!_S}}$ is the velocity at the drop surface and $D_{\!_S}$ denotes the surface-diffusion coefficient.

We adopt the following non-dimensional scheme: length $\sim$ $a$; electric potential $\sim$ $aE$; conductivity $\sim$ $\sigma_e$; permittivity $\sim$ $\epsilon_e$ and viscosity $\sim$ $\mu_e$; velocity $\sim$ $ \widetilde U_{\!_{HR}}$; surfactant concentration $\sim$ $ \widetilde{\Gamma}_{eq} $; hydrodynamic stress $\sim$ $\mu_e \widetilde U_{\!_{HR}}/{a}$ and electric stress $\sim$ $\epsilon_eE^2$. The choice of velocity scale as the Hadamard-Rybczynski velocity ($ \widetilde U_{\!_{HR}}$) over the eletrohydrodynamic velocity scale ($\epsilon_eaE^2/\mu_e$) is adopted with a rationale
 to obtain a correction of settling velocity \citep{Xu2006,Mandal2017,Mandal2017a} in the following analysis.
Henceforth we will also use the following notations for various parameters to describe the physical situation:
viscosity ratio $\lambda=\mu_i/\mu_e$,
conductivity ratio $R=\sigma_i/\sigma_e$,
permittivity ratio $S=\epsilon_i/\epsilon_e$,
electric Reynolds number $Re_{\!_E}=\epsilon_e \widetilde U_{\!_{HR}}/a\sigma_e$,
Capillary number $Ca=\mu_e \widetilde U_{\!_{HR}}/\gamma_{eq}$, Mason number $ M=\epsilon_eaE^2/\mu_e \widetilde U_{\!_{HR}} $ and surface P\'eclet number $ Pe=a\,\widetilde U_{\!_{HR}}/D_{\!_S} $. Now onwards we will drop all the $\,\,\widetilde{}\,\,$  symbols in the dimensional variables  to denote the corresponding dimensionless forms for brevity.

\subsubsection{Electrostatic problem}
\label{ssec:E_prob}

According to the leaky dielectric model introduced by \citet{Melcher1969} the electric potential ($\varphi$) satisfies the Laplace equation as follows \citep{Melcher1969,Saville1997,Taylor1966,Arp1980}:

\begin{equation}
 \begin{aligned}
\nabla^2\varphi_i&=0  \\
\nabla^2\varphi_e&=0  
 \end{aligned}
\label{eq:laplace}
\end{equation}
The electric potential is subjected to the following boundary conditions:
\begin{itemize}
	\item[] (i)\,\,\,\,$\varphi_i$ is bounded at $r=0$.
	\item[] (ii)\,\,\,\,$\varphi_e$ satisfies the far-field condition: 
	as $r\to\infty$, \; $\varphi_e=-rP_1(\eta)$, where $ \eta=\cos{\theta} $ and $ P_1 $ is the Legendre polynomial of first degree.
	\item[] (iii)\,\,\,\,At the drop interface the potential is continuous, i.e. at $r=r_{\!_S}(\theta)$, \; $\varphi_i=\varphi_e$.
	\item[] (iv)\,\,\,\,The interfacial charge conservation at the steady state is satisfied by the electric potential i.e., 
	\begin{equation}
\textrm{\\at} \,\, r=r_{\!_S}(\theta)\quad \mathbf{n}\cdot (R\nabla \varphi_i-\nabla\varphi_e)=-Re_{\!_E}\nabla_{\!_S}\cdot(q\,_{\!_S}\mathbf{u}_{\!_S}), 
\label{eq:cc}
	\end{equation}
	 where $\mathbf{n}$ represents the outward unit normal at the drop interface, $\nabla_{\!_S}$ is the surface divergence operator and $\mathbf{u}_{\!_S}$ is the fluid velocity at the interface. Here $q_{\!_{\,S}}$ represents the interfacial charge density defined as $q_{\!_{\,S}}=\mathbf{n}\cdot (S\nabla \varphi_i-\nabla\varphi_e)\bigg|_{r=r_{\!_S}}^{}$.
	  In the above equation (\eqref{eq:cc}), the left hand side represents current due to Ohmic conduction while the right hand side stands for the current due to charge convection at the droplet surface. The coupling between the fluid flow and electric field is clear from the above boundary conditions.

\end{itemize}
\subsubsection{Hydrodynamic problem}
\label{ssec:H_prob}
In the creeping flow limit, the velocity ($ \mathbf{u}_{\!\,_i},\mathbf{u}_{\!\,_e} $) and pressure ($p_{\!\,_i},p_{\!\,_e}$) both inside and outside the drop satisfy the Stokes equation as: 
\begin{equation}
\begin{aligned}
\nabla{p}_i&=\lambda\nabla^2\mathbf{u}_i, &&\nabla\cdot{\mathbf{u}_i}=0\\
\nabla{p}_e&=\,\,\,\,\nabla^2\mathbf{u}_e,        &&\nabla\cdot{\mathbf{u}_e}=0
\end{aligned}
\label{eq:momentum}
\end{equation}
The corresponding velocity components can be expressed in terms of stream function ($\Psi$) as $u_r=-\dfrac{1}{r^2}\dfrac{\partial \Psi}{\partial \eta}$ and $u_\theta=-\dfrac{1}{r\sqrt{1-{\eta}^2}}\dfrac{\partial \Psi}{\partial r}$.
Under these considerations, \eqref{eq:momentum} becomes a set of fourth order partial differential equations, given as
\begin{equation}
\begin{aligned}
\mathcal{L}^2(\mathcal{L}^2\Psi_i&=0)  \\
\mathcal{L}^2(\mathcal{L}^2\Psi_e&=0)\,,  
\end{aligned}
\label{eq:stokes}
\end{equation}
where $\Psi_i\,$, $\Psi_e$ are the streamfunctions for the drop phase and the continuous phase, respectively and $\mathcal{L}^2$ is a linear operator 
in spherical polar co-ordinate system \citep{Leal2007}.

The velocity and pressure fields of the drop phase ($\mathbf{{u}_i},p_i$) and that of the surrounding medium ($\mathbf{{u}_e},p_e$) satisfy the following boundary conditions:
\begin{itemize}
	\item[] (i)\,\,\,\, $\mathbf{u_i},p_i$ are bounded inside the drop.
	\item[] (ii)\,\,\,\, $\mathbf{u_e},p_e$ are bounded outside the drop and the far field condition of uniform streaming flow, at  $r\to\infty$, \; $\mathbf{u}_e=-\mathbf{U}_d$, is obeyed by the outside velocity.
	\item[] (iii)\,\,\,\,At the drop interface the velocity is continuous i.e. $\mathbf{u}_i=\mathbf{u}_e$ and at steady state normal component of velocity vanishes ($\mathbf{u}_i\cdot{\mathbf{n}}=\mathbf{u}_e\cdot{\mathbf{n}}=0$).
	\item[] (iv)\,\,\,\,   Gradient in surface tension generated due nonuniformity in surfactant concentration gives rise to  an additional component of interfacial stress known as the Marangoni stress ($\boldsymbol{T}^{Ma}$).  The dimensionless form of $\boldsymbol{T}^{Ma}$ is given by
	\begin{equation}
\boldsymbol{T}^{Ma}=Ma\,[(1-\Gamma)(\nabla\cdot{\mathbf{n}})\mathbf{n}+\nabla_{\!_S}\Gamma],
	\end{equation}
where the Marangoni number ($ Ma $) is defined as $ Ma=R_gT\,\widetilde{\Gamma}_{eq}/\mu_e \widetilde U_{HR} $. This embodies the relative strength of Marangoni stress and the viscous stress.
	The interfacial balance between the  hydrodynamic and electrical components of stress, which would exist otherwise, now gets modified by the appearance of        $\boldsymbol{T}^{Ma}$. The resulting stress balance equation at the interface takes the following form
	\begin{equation}
	\\\textrm{at } \,r=r_{\!_S}(\theta),\quad \llbracket\boldsymbol{T}^H\rrbracket+M\:\llbracket\boldsymbol{T}^E\rrbracket=\frac{1}{Ca}(\nabla\cdot{\mathbf{n}})\mathbf{n}+Ma\,[(1-\Gamma)(\nabla\cdot{\mathbf{n}})\mathbf{n}+\nabla_{\!_S}\Gamma]. 
	\label{eq:stress_bal}
	\end{equation}
	Here $\llbracket\,\,\rrbracket$ represents the jump or difference of any quantity (e.g.\ $\llbracket\boldsymbol{T}\rrbracket=\boldsymbol{T}_e-\boldsymbol{T}_i$) across the interface, while $\boldsymbol{T}^H=
	\boldsymbol{\tau}^H\cdot{\mathbf{n}}$ and $\boldsymbol{T}^E=
	\boldsymbol{\tau}^E\cdot{\mathbf{n}}$ are the viscous and electric parts of traction vector. The forms of viscous and electric stress tensors are given by
    \,$\boldsymbol{\tau}^H_i=-p_i\mathbf{I}+{\lambda}[\nabla\mathbf{u}_i+(\nabla\mathbf{u}_i)^T]$; \,$\boldsymbol{\tau}^H_e=-p_e\mathbf{I}+[\nabla\mathbf{u}_e+(\nabla\mathbf{u}_e)^T]$;\,$\boldsymbol{\tau}^E_i=S\left[\mathbf{E}_i\mathbf{E}^T_i-\frac{1}{2}\left|{\mathbf{E}_i}\right|^2\mathbf{I}\right]$ and $\boldsymbol{\tau}^E_e=\left[\mathbf{E}_e\mathbf{E}^T_e-\frac{1}{2}\left|{\mathbf{E}_e}\right|^2\mathbf{I}\right]$.

\end{itemize}

 For later convenience $ Ma $ is related to the physicochemical parameter $ \beta $ by the relation $Ma=\dfrac{\beta}{Ca(1-\beta)}$, where
$ \beta $ is known as the surface elasticity number and defined as 
$\beta=
-\dfrac{d(\widetilde{\gamma}/\widetilde\gamma_c)}
{d(\widetilde\Gamma / \widetilde\Gamma_{eq})}
=\dfrac{R_gT\,\widetilde\Gamma_{eq}}{\widetilde\gamma_c}$. This quantifies the sensitivity of the interfacial tension on the surfactant concentration. Theoretically $ \beta $ can vary between 0 and 1 \citep{Stone1990,Li1997}. To visualize the importance of the specific terms in the right hand side of \eqref{eq:stress_bal},  tangential and normal components of stress balance  are presented separately as:
\begin{equation}
\textrm{\\at } \,r=r_{\!_S}(\theta),\quad 
\llbracket\boldsymbol{T}^H_t\rrbracket+M\:\llbracket\boldsymbol{T}^E_t\rrbracket=\underbrace{\frac{\beta}{Ca(1-\beta)}\,(\nabla_{\!_S}\Gamma)\cdot \mathbf{t}}_{\textrm{ Marangoni effect}}
\label{eq:t_stress_bal}\,,
\end{equation}
and
\begin{equation}
\textrm{\\at} \,r=r_{\!_S}(\theta),\quad \llbracket\boldsymbol{T}^H_n\rrbracket+M\:\llbracket\boldsymbol{T}^E_n\rrbracket=\frac{1}{Ca}(\nabla\cdot{\mathbf{n}})+\underbrace{\frac{\beta}{Ca(1-\beta)}\,[(1-\Gamma)(\nabla\cdot{\mathbf{n}})]}_\text{Marangoni effect}\,,
\label{eq:n_stress_bal} 
\end{equation}
where $ \mathbf{t} $ is the unit vector in the tangential direction at the drop surface.

The dimensionless form of the surfactant transport equation becomes
\begin{equation}
Pe\nabla_S\cdot({\mathbf{u}_{\!_{\,S}}}\Gamma)=\nabla^2_{\!_S}\Gamma
\label{eq:surf_transport}
\end{equation}
In addition, the surfactant present on the drop surface must satisfy a mass conservation constraint \citep{Kim1989} given by
\begin{equation}
\label{mass-constraint}
 \int_{0}^{\pi} \Gamma(\theta)\, r^2_{\!_S}(\theta)\sin{\theta} \,d\theta=2 
\end{equation}
We note that the mathematical model becomes non-linear in nature owing to the presence of convective transport at the drop interface as depicted by the left hand side of \eqref{eq:surf_transport}. Secondly the flow field and the surfactant distribution at the interface become coupled to each other. 

	\subsection{Expansion in terms of $ Re_{\!_E} $ and $Ca$} \label{ss:perturbation}
	
Source of complexity in solving the governing equations and boundary conditions arises from the  convection-diffusion transport equation governing the distribution of insoluble surfactants               (\eqref{eq:surf_transport}) on the surface and its coupling with the flow problem (\eqref{eq:stress_bal}). Moreover, the deformed surface is a priori unknown and must be found out through normal stress balance.  Adding to these, is the nonlinearity associated with the charge convection equation, thus rendering the physical system at hand impossible to solve analytically for arbitrary range of parameters. Hence to proceed with an analytical treatment we evaluate various dimensionless parameters based on the  dimensional values of the property and geometric dimensions reported elsewhere \citep{Xu2006,Mhatre2013}. 
We  then employ a double asymptotic expansion with $Ca$ and $Re_E$ as perturbation parameters \citep{Xu2006,Mandal2016}. Accordingly any generic variable, $\xi$, can be expanded in the following form
 \begin{equation}
 \xi=\xi^{(0)}+Ca\,\xi^{(Ca)}+Re_{\!_E}\,\xi^{(Re_{\!_E})}+O(Re_{\!_E}^2, Ca^2, CaRe_{\!_E})
 \label{eq:general_perturb}
 \end{equation}
 Special care is to be taken while expanding the pressure and stress inside the droplet. In order to balance the capillary pressure in absence of any flow field, these are expanded as \citep{Chan1979,Mandal2016}:
 \begin{equation}
 \label{eq:cap-press}
 \left.
 \begin{split}
 p_i &=\frac{1}{Ca}p^{(1/Ca)}_i+p^{(0)}_i+Ca\,p^{(Ca)}_i+Re_{\!_E}\,p^{(Re_{\!_E})}_i+\ldots\\
\boldsymbol{\tau_i} &=\frac{1}{Ca}\boldsymbol{\tau_i}^{(1/Ca)}+\boldsymbol{\tau_i}^{(0)}+Ca\,\boldsymbol{\tau_i}^{(Ca)}+Re_{\!_E}\,\boldsymbol{\tau_i}^{(Re_{\!_E})}+\ldots
 \end{split}
 \right \}
 \end{equation}

 The drop shape is unknown a priori and we seek for a solution of the same by expanding it in the form:
 \begin{equation}
 r_{\!_S}=1+Ca\,f^{(Ca)}+CaRe_{\!_E}\,f^{(CaRe_{\!_E})}+Ca^2\,f^{(Ca^2)}+\ldots
 \label{eq:shape_perturb}
 \end{equation} Here $f^{(Ca)}$, $f^{(CaRe_E)}$ and $f^{(Ca^2)}$ are the shape functions which can be expressed as a sum of surface harmonics:
 \begin{equation}\label{eq:f_expand}
 f^{(j)}=\sum_{n=1}^{\infty} L^{(j)}_nP_n(\eta); \quad j=Ca,\,CaRe_{\!_E},\,Ca^2
 \end{equation}

We define the drop deformation parameter as
 
 \begin{equation} \label{eq:deformation-def}
 \mathcal{D}=\dfrac{r_{\!_S}\big|_{\theta=0}-r_{\!_S}\big|_{\theta=\pi/2}}{r_{\!_S}\big|_{\theta=0}+r_{\!_S}\big|_{\theta=\pi/2}},
 \end{equation} which will be used later in the interpretation of different deformation-related effects.
 
 The surfactant concentration is also expected to deviate from the equilibrium value owing to deformation of the drop from the initial sphericity. Thus the concentration ($ \Gamma $) is also expanded in a similar fashion to that of the shape deformation \citep{Ha1995,Stone1990,Mandal2016b} and given as 
 \begin{equation}
 \Gamma=\Gamma^{(0)}+Ca\,\Gamma^{(Ca)}+CaRe_{\!_E}\,\Gamma^{(CaRe_{\!_E})}+Ca^2\Gamma^{(Ca^2)}+\ldots
 \label{eq:surf_perturb}
 \end{equation} Here $ \Gamma^{(0)}$ turns out to be 1 when the mass conservation constraint (\eqref{mass-constraint}) is incorporated. Similar to the shape function, the remaining surfactant harmonics can be expressed as 
 \begin{equation}\label{eq:g_expand}
 \Gamma^{(j)}=\sum_{n=1}^{\infty} \Gamma^{(j)}_nP_n(\eta); \quad j=Ca,\,CaRe_{\!_E},\,Ca^2
 \end{equation}
 
In what follows we will proceed to obtain analytical solutions for the two different regimes: (i) diffusion dominated regime  or the low P\'eclet limit ($Pe\ll1$) and (ii) convection dominated regime or the high P\'eclet regime ($Pe\gg1$). The practical relevance of these regimes are discussed in great detail by \citet{Das2017a} with reference to the experimental situations \citep{Stebe1991}. It is to be noted that having chosen $ Ca $ as an independent dimensionless parameter, we cannot consider $Pe$ as another independent parameter \citep{Stone1990,Mandal2016a}, since both these dimensionless numbers contain the same $\widetilde{U}_{\!_{HR}}$ as the velocity scale. Instead we use a ratio of the two, $k=\dfrac{Pe}{Ca}=\dfrac{\widetilde{\Gamma}_{eq}\,a}{\mu_e\,D_{\!_S}}$ which becomes a $O(1)$ dimensionless constant for $ Pe\ll1 $and consists of various material properties of surface-active agents.

The detailed expressions for the governing equations and boundary conditions of the electric as well as flow problem for the case of a clean, surfactant-free drop can be found in \citet{Mandal2017,Mandal2017a}. The boundary conditions of the electrostatic problem and the velocity boundary conditions (no slip and kinematic condition),  remaining unaltered, will not be repeated here. However the presence of surface-active molecules on the drop interface triggers Marangoni flow owing to the variability in interfacial tension. This in turn couples the surfactant transport phenomena with the flow problem through the Marangoni stress term in the stress balance equation (\eqref{eq:stress_bal}). The tangential and normal stress balance equations along with the corresponding equations for surfactant transport, at different perturbation orders, are discussed below.
\\
\\\textbf{Leading-order:}\\

The boundary conditions, in this order, are  evaluated on the undeformed surface (at $r=1$) and are as follows:\\ 
\\
(i) \textit{Tangential stress balance  }: 
 	\begin{equation}
\textrm{\\at} \,r=1,\quad \llbracket\boldsymbol{T}^{H(0)}_{\theta}\rrbracket+M\:\llbracket\boldsymbol{T}^{E(0)}_{\theta}\rrbracket=\frac{\beta}{1-\beta}\frac{d\,\Gamma}{d\theta}^{(Ca)} 
 \label{eq:leading-order-tangential}
 \end{equation}
(ii) \textit{Normal stress balance  }: 
\begin{equation}
\textrm{\\at} \,r=1,\quad \llbracket\boldsymbol{T}^{H(0)}_{r}\rrbracket+M\:\llbracket\boldsymbol{T}^{E(0)}_{r}\rrbracket=-(2f^{(Ca)}+\nabla^2f^{(Ca)})-\frac{2\beta}{1-\beta}\Gamma^{(Ca)}
\label{eq:leading-order-normal}
\end{equation}
(iii) \textit{Surfactant transport  }: 
\begin{equation}
\begin{aligned}
&\textrm{Low P\'eclet number limit:} &&\textrm{\qquad at}  \,\,r=1,\quad 
k\nabla_{\!_S}\cdot({\mathbf{u}^{(0)}_{\!_S}})=\nabla^2_{\!_S}\Gamma^{(Ca)}\\
&\textrm{High P\'eclet number limit:}&&\textrm{\qquad at} \,\,r=1,\quad 
\nabla_{\!_S}\cdot({\mathbf{u}^{(0)}_{\!_S}}) \, \,\,=0
\end{aligned}
\label{eq:leading-order-surf}
\end{equation}
\\
\\ $\bf{O(Re_{\!_E}):}$ \textbf{Effect of finite charge convection}\\

Now we proceed to capture the effects of surface charge convection by adopting electric Reynolds number ($ Re_{\!_E}$) as a perturbation parameter without taking into account of the shape distortion. Thus the boundary conditions are applied at $r=1$ as done for the leading-order case. The corresponding equations take the forms:\\
\\
(i) \textit{Tangential stress balance  }: 
\begin{equation}
\textrm{at}  \,r=1,\quad \llbracket\boldsymbol{T}^{H(Re_{\!_E})}_{t}\rrbracket+M\:\llbracket\boldsymbol{T}^{E(Re_{\!_E})}_{t}\rrbracket=\frac{\beta}{1-\beta}\frac{d\,\Gamma}{d\theta}^{(CaRe_{\!_E})} 
\label{eq:ReE-order-tangential}
\end{equation}
(ii) \textit{Normal stress balance  }: 
\begin{equation}
\textrm{at} \,r=1,\quad \llbracket\boldsymbol{T}^{H(Re_{\!_E})}_{n}\rrbracket+M\:\llbracket\boldsymbol{T}^{E(Re_{\!_E})}_{n}\rrbracket=-(2f^{(CaRe_{\!_E})}+\nabla^2f^{(CaRe_{\!_E})})-\frac{2\beta}{1-\beta}\Gamma^{(CaRe_{\!_E})}
\label{eq:ReE-order-normal}
\end{equation}
(iii) \textit{Surfactant transport  }: 
\begin{equation}
\begin{aligned}
&\textrm{Low P\'eclet number limit:\,\qquad at} \,\,r=1,\quad 
k\nabla_{\!_S}\cdot({\mathbf{u}^{(Re_{\!_E})}_{\!_S}})&&=\nabla^2_{\!_S}\Gamma^{(CaRe_{\!_E})}\\
&\textrm{High P\'eclet number limit:\qquad at} \,\,r=1,\quad 
\nabla_{\!_S}\cdot({\mathbf{u}^{(Re_{\!_E})}_{\!_S}})&&=0
\end{aligned}
\label{eq:ReE-order-surf}
\end{equation}\\
\\ {$\mathbf{O(Ca):}$ \textbf{Effect of drop-shape deformation}}\\

All the quantities at this order are to be evaluated on the deformed interface $ r=r_{\!_S}(\theta)=1+Ca\,f^{(Ca)}$ which is calculated with the help of leading-order solution for shape function $f^{(Ca)} $ . However, since the drop shape is not known a priori, following the domain perturbation technique, we evaluate the corresponding quantities at $r=1$ by a Taylor series expansion about $r=1$      \citep{Happel1983,Bandopadhyay2016,Mandal2016}. Mathematically, for a generic quantity this can be stated as 
\begin{equation}\label{eq:taylor-expansion}
\left[\xi\big|_{ \\at\,r=1+Ca\,f^{(Ca)} }^{}\right]^{(Ca)}={\xi}^{(Ca)}\big|_{ \\at\,r=1 }^{}+f^{(Ca)}\frac{\partial\xi^{(0)}}{\partial r}\bigg|_{r=1}^{}
\end{equation}
The relevant stress boundary conditions and surfactant transport equations take the following forms:\\
\\            
(i) \textit{Tangential stress balance  }: 
\begin{equation}
\textrm{at} \,r=1,\quad \llbracket\boldsymbol{T}^{H(Ca)}_{t}\rrbracket+M\:\llbracket\boldsymbol{T}^{E(Ca)}_{t}\rrbracket=\frac{\beta}{1-\beta}\left[\frac{d\,\Gamma}{d\theta}^{(Ca^2)}+f^{Ca}\frac{\partial}{\partial r}\left(    \frac{1}{r}\frac{\partial \Gamma}{\partial \theta}^{(Ca^2)}\right)\right] 
\label{eq:Ca-order-tangential}
\end{equation}
(ii) \textit{Normal stress balance  }: 
\begin{equation}
\begin{aligned}
\textrm{at} \,r=1,\quad \llbracket\boldsymbol{T}^{H(Ca)}_{n}\rrbracket+M\:\llbracket\boldsymbol{T}^{E(Ca)}_{n}\rrbracket=2f^{(Ca)}(f^{(Ca)}+\nabla^2f^{(Ca)})-(2f^{(Ca^2)}+\nabla^2f^{(Ca^2)})\\
-\frac{2\beta}{1-\beta}\Gamma^{(Ca^2)}+\frac{\beta}{1-\beta}\Gamma^{(Ca)} (2f^{(Ca)}+\nabla^2f^{(Ca)})
\end{aligned}
\label{eq:Ca-order-normal}
\end{equation}
(iii) \textit{Surfactant transport  }: 
\begin{equation}
\begin{split}
&\textrm{Low P\'eclet number limit:}\\
&\qquad \textrm{at} \,\,r=1,\quad 
k\nabla_{\!_S}\cdot({\mathbf{u}^{(0)}_{\!_S}}\Gamma^{(Ca)}+{\mathbf{u}^{(Ca)}_{\!_S}})=\nabla^2_{\!_S}\Gamma^{(Ca^2)}+f^{(Ca)}\frac{\partial}{\partial r}\left(\nabla^2_{\!_S}\Gamma^{(Ca)}\right)\\
&\qquad \qquad \qquad \qquad +\frac{\partial f}{\partial \theta}^{(Ca)}
\left(\frac{2}{r^2}\frac{\partial \Gamma}{\partial \theta}^{(Ca)} +\frac{\partial}{\partial r}\left(  \frac{1}{r}\frac{\partial \Gamma}{\partial \theta}^{(Ca)}\right)-\frac{1}{r^2}\frac{\partial \Gamma}{\partial \theta}^{(Ca)}\right) \\
&\textrm{High P\'eclet number limit:}\quad \\
&\qquad \textrm{at} \,\,r=1,\qquad
\nabla_{\!_S}\cdot({\mathbf{u}^{(0)}_{\!_S}}\Gamma^{(Ca)}+{\mathbf{u}^{(Ca)}_{\!_S}})=0
%\end{aligned}
\end{split}
\label{eq:Ca-order-surf}
\end{equation}
The expressions for hydrodynamic and electric traction vectors as appeared in \eqref{eq:Ca-order-tangential} and \eqref{eq:Ca-order-normal} can be found elsewhere \citep{Mandal2016b} and not repeated here.

\subsection{Solution methodology}
 \label{ss:solution}
In this section we describe the solution methodology to obtain the detailed expressions of electric potential, stream function, shape deformation and the surfactant distribution at different order of perturbations. Since at each order the electric potential satisfies the Laplace equation, the general solution of  $ \varphi_i $ and $ \varphi_e $ can be expressed as a sum of spherical harmonics, keeping in view of the boundedness and far-field conditions:

\begin{equation}
\begin{aligned}
&\varphi^{(j)}_i=\sum_{n=0}^{\infty} r^{n}a^{(j)}_nP_n(\eta)  \\
&\varphi^{(j)}_e=\varphi^{(j)}_\infty + \sum_{n=0}^{\infty} r^{-n-1}b^{(j)}_{-n-1}P_n(\eta) ,
\end{aligned}
\label{eq:sol_pot}
\end{equation}
where $\varphi^{(j)}_\infty$ denotes the unperturbed potential to match with the far-field condition of uniform electric field and only exists in the leading-order. Thus it takes a form of $ \varphi^{(0)}_\infty=-rP_1(\eta) $.
Similar to the electric potential, the general solution of  stream-function which satisfies the fourth-order partial differential equation in both the regions, takes the form:

\begin{equation}
\begin{aligned}
&\Psi^{(j)}_i=\sum_{n=1}^{\infty} \left[A^{(j)}_nr^{n+3}+B^{(j)}_nr^{n+1} \right]Q_n(\eta)  \\
&\Psi^{(j)}_e=U^{(j)}_dr^2Q_1(\eta) + \sum_{n=1}^{\infty} \left[C^{(j)}_nr^{2-n}+D^{(j)}_nr^{-n} \right]Q_n(\eta) 
\end{aligned}
\label{eq:sol_sf}
\end{equation}
Here $ Q_n(\eta)=\int_{-1}^{\eta} P_n(\eta)d\eta $ denotes the $ n $-th degree of Gegenbauer polynomial \citep{Leal2007}. Now the task remains to obtain the coefficients of the electric potential, stream function and surfactant concentration using appropriate set of boundary conditions at each order of perturbation and invoking the proper orthogonality conditions of the Legendre and Gegenbauer polynomials.

In order to obtain the drop migration velocity, at different orders of perturbation, we follow the steps as followed by several authors previously \citep{Xu2006,Mandal2017}. In the first step the constant coefficients appearing in the electric potential distribution ($a_n,b_{-n-1} $) (\eqref{eq:sol_pot}), are obtained by solving the boundary conditions as mentioned in section (\ref{ssec:E_prob}). The solution is then used to determine the electric force on the drop given by $ \boldsymbol{F}^E=2\pi\int_{\theta=0}^{\pi} \left(\boldsymbol{\tau}^E_e\cdot\mathbf{n} \right)r^2_{\!_S} \sin{\theta}\,d\theta  $. The next step is to determine the stream function coefficients ($A_n,B_n,C_n,D_n$)(\eqref{eq:sol_sf}) to obtain hydrodynamic force on the drop by using the expression 
 $ \boldsymbol{F}^H=2\pi\int_{\theta=0}^{\pi} \left(\boldsymbol{\tau}^H_e\cdot\mathbf{n} \right)r^2_{\!_S} \sin{\theta}\,d\theta  $.
 Consequently, the fact that the drop is force-free yields 
 \begin{equation}\label{eq:force-free}
 \boldsymbol{F}^H+\boldsymbol{F}^E+\boldsymbol{F}^B=\boldsymbol{0},
 \end{equation}
 allowing $ U_d $ to be determined. Here $\boldsymbol{F}^B=\dfrac{2\pi(2\pi+3\lambda)}{1+\lambda}\,\mathbf{e}_z$ is the buoyancy force acting on the drop and appears in the leading-order force-balance only. Moreover, the  hydrodynamic force on the drop can be conceptually split into two parts, namely the drag force experienced by the falling drop in a viscous fluid ($ \boldsymbol{F}^{H}_{drag} $) and the electrohydrodynamic force ($ \boldsymbol{F}^{H}_{EHD} $). The electrohydrodynamic force owes its existence to the imbalance of tangential stress at the interface and the consequent asymmetric flow.  
 To bring out the alterations in the flow field and potential distribution, induced by nonuniform distribution of surfactant molecules on the drop surface, the surfactant transport equation has to be solved concurrently with the velocity and stress boundary conditions.  
 
 Before proceeding with the explicit expressions of the solution of various order of perturbation, we require the outward unit normal ($\mathbf{n}$) and the curvature of the drop surface ($\nabla\cdot\mathbf{n} $) at any point on the interface which are calculated as follows \citep{Ramachandran2012}:
 \begin{equation}\label{eq:unit-normal}
\begin{split}
\mathbf{n}= \frac{\nabla(r-r_{\!_S})}{|\nabla(r-r_{\!_S})|}
=&\mathbf{e}_r-Ca\nabla f^{(Ca)}-CaRe_{\!_E}\nabla f^{(CaRe_{\!_E})} \\
& -Ca^2 \left\{\nabla f^{(Ca^2)}+\frac{1}{2} \left( \nabla f^{(Ca)} \cdot \nabla f^{(Ca)}   \right)\mathbf{e}_r
 \right\} -\ldots
 \end{split}
 \end{equation}
 \begin{equation}\label{eq:curvature}
\begin{split}
\nabla \cdot \mathbf{n}= 2-
Ca\left(2f^{(Ca)}+\nabla^2f^{(Ca)} \right)
-CaRe_{\!_E}\left(2f^{(CaRe_{\!_E})}+\nabla^2f^{(CaRe_{\!_E})} \right) \\
 -Ca^2 \left\{-2f^{(Ca)}\left(f^{(Ca)}+\nabla^2 f^{(Ca)} \right)
 +2f^{(Ca^2)}+\nabla^2 f^{(Ca^2)}
\right\} -\ldots
\end{split}
\end{equation}
\subsubsection{Leading order solution}
\label{sss:O(1)_solution}
The leading-order electric potential distribution does not contain any information about the sedimentation effects, charge convection or the surfactant contribution. It remains the same as the solution found in the pioneering work of  \citet{Taylor1966}:
\begin{equation}
\label{eq:leading-pot}
\left.
\begin{split}
\varphi^{(0)}_i & =-{\frac {3r\cos {\theta} }{R+2}}\\
\varphi^{(0)}_e & =-r\cos {\theta} +{\frac { \left( R-1 \right) \cos {\theta} }{ \left( R+2 \right) {r}^{2}}}
\end{split}
\right \}
\end{equation}
The corresponding charge distribution is given by 
\begin{equation}\label{eq:cd-leading}
q^{(0)}_{\!_{\,S}}=\frac {3\cos{\theta} \left( R-S \right) }{R+2}
\end{equation}

To obtain the pressure and velocity fields inside and outside of the droplet we need the expressions for streamfunction in the corresponding case as given below:
\begin{equation}
\label{eq:leading-sf}
\left.
\begin{split}
\Psi^{(0)}_i & =\sum_{n=1}^{2}\left( A^{(0)}_{{n}}{r}^{n+3}+B^{(0)}_{{n}}{r}^{n+1} \right) Q_{{n}} \left( \eta\right) \\
\Psi^{(0)}_e & ={r}^{2}U^{(0)}_dQ_1(\eta)+\sum_{n=1}^{2}\left( C^{(0)}_{{n}}{r}^{2-n}+D^{(0)}_{{n}}{r}^{-n} \right) Q_{{n}} \left( \eta\right) \\
\end{split}
\right \},
\end{equation}
where the constants $A^{(0)}_n,B^{(0)}_n,C^{(0)}_n,D^{(0)}_n$ can be found  in the supplementary MATLAB file. Though at each order of perturbation, coefficients of the streamfunction are different for low and high P\'eclet number limits,  the high P\'eclet number limit constants can be obtained by taking a limit $k\to\infty$ of their low P\'eclet  number limit counterparts \citep{Mandal2016b}. Subsequently the inner region streamfunction becomes identically zero in $k\to\infty$ limit rendering the surface velocity to vanish in this case. 

The corresponding local surfactant concentration distribution is obtained as:
\begin{equation}\label{eq:surf-Ca}
\Gamma^{(Ca)}=\sum_{n=1}^{2} \Gamma^{(Ca)}_nP_n(\eta),
\end{equation}where 
\begin{equation}
\label{eq:leading-st}
\left.
\begin{split}
\Gamma^{(Ca)}_1 & ={\frac {k \left( 2+3\,\lambda \right)  \left( 1-\beta\right) }{2 \left( 1+\lambda \right)  \left(  \left( 3\,\lambda-k+2 \right) \beta-3\,\lambda-2 \right) }}\\
\Gamma^{(Ca)}_2 &  =\left(
\frac
{
	-3kM   \left( R-S \right) (1-\beta) 
}
{
	\splitfrac
	{
\left(  \left( -{R}^{2}-4\,R-4 \right) k+5\,\lambda\,{R}^{2}+5
		\,{R}^{2}+20\,\lambda\,R+20\,R+20\,\lambda+20 \right) \beta
	}
	{
-5\,\lambda
		\,{R}^{2}-5\,{R}^{2}-20\,\lambda\,R-20\,R-20\,\lambda-20
	}
}
\right).
\end{split}
\right \}
\end{equation}

Upon using the force free condition (\eqref{eq:force-free}), the leading-order drop velocity is obtained as
\begin{subequations}
	\begin{equation}\label{eq:Ud-leading-low}
U^{(0)}_d=1-
\underbrace{\frac{1}{3}\,
	{\frac 
		{k\,\beta}
		{ 
			\left( 1+\lambda \right)  
			 \left((2+3\lambda)(1-\beta)+k\beta\right)}}
		}
		_\text{correction due to surfactant}
	\end{equation}
In the High P\'eclet number limit the same expression becomes:
\begin{equation}\label{eq:Ud-leading-high}
U^{(0)}_d=1-\underbrace{\frac{1}{3(1+\lambda)}}_{\substack{\text{correction due} \\ \text{to surfactant}}}
\end{equation}
\end{subequations}
\\Interestingly, although in the present problem both the settling and EHD effects are considered, the leading-order droplet velocity does not contain any electric effects and the leading  order correction to the Hadamard - Rybczynski velocity ($\widetilde{U}_{\!_{HR}}$) comes solely from surfactant effects. 
Such a condition is realized since in the  leading-order the electrohydrodynamic force ($ \boldsymbol{F}^{H}_{EHD} $) does not contribute to the hydrodynamic force and a balance is established only between the viscous drag  ($ \boldsymbol{F}^{H}_{drag} $)  and buoyancy ($\boldsymbol{F}^B$).  
In addition, the leading-order drop velocity always gets reduced if the surfactant induced interfacial tension variation is present, since the correction term comes out to be positive always
$ \left((2+3\lambda)(1-\beta)+k\beta\right)>0$ considering the practical range of the physical parameters $(k\sim O(1), \,0<\beta<1,\, \lambda>0)$.

Using the information about the surfactant distribution and drop velocity, the normal stress balance equation (\eqref{eq:leading-order-normal}) is solved to obtain the harmonics of shape function. The shape function thus obtained takes the form $f^{(Ca)}= L^{(Ca)}_2P_2(\eta)$, where
\begin{subequations}
\begin{equation}\label{eq:sol-leading-shape}
L^{(Ca)}_2=\frac{3}{4}\,{\frac {M \left(  c_1\beta k+c_2 \beta+c_3
		 \right) }
	{\left(  d_1\beta k+d_2 \beta+d_3
		\right)}} \quad \text{in low P\'eclet limit and}
\end{equation}
\begin{equation}\label{eq:high-leading-deform}
L^{(Ca)}_2=\frac{3}{4}\,{\frac { M\left( {R}^{2}+2\,R-4\,S+1 \right) }{{R}^{2}+4\,R+4}} \quad \text{in high P\'eclet limit},
\end{equation}
\end{subequations}
\\ where $c_1=\left( -{R}^{2}-2\,R+4\,S-1 \right) $,
 $c_2=\left( 5\,\lambda\,{R}^{2}+5\,{R}^{2}+9\,\lambda\,R-19\,S\lambda+6\,R-16\,S+5\,\lambda+5 \right)$,
$c_3=-c_2$, $d_1=-{R}^{2}-4\,R-4$, $d_2=5\,\lambda\,{R}^{2}+5\,{R}^{2}+20\,\lambda\,R+20\,R+20\,\lambda+20$ and $d_3=-d_2$.
\subsubsection{$O(Re_{\!_E})$ solution}
\label{sss:O(ReE)_solution}
Using the solution of leading-order as obtained above and following the boundary conditions in section \ref{ss:perturbation}, the electrostatic potential is calculated as follows 
\begin{equation}
\label{eq:sol_pot-ReE}
\left.
\begin{split}
\varphi^{(Re_{\!_E})}_i & =\sum_{n=1}^{3} r^{n}a^{(Re_{\!_E})}_nP_n(\eta) \\
\varphi^{(Re_{\!_E})}_e & = \sum_{n=1}^{3} r^{-n-1}b^{(Re_{\!_E})}_{-n-1}P_n(\eta) , \\
\end{split}
\right \}
\end{equation}
where the constants $a^{(Re_{\!_E})}_n, b^{(Re_{\!_E})}_{-n-1}$ are provided in the supplementary MATLAB file.The corresponding surface charge density turns out to be \\
\begin{equation}\label{eq:sol-sc-ReE}
q^{(Re_{\!_E})}_{\,\!_S}=\frac{108\,\left(1-\beta\right)\,\left(R-S\right)}{5\,d_4}\left[c_4\cos^3{\theta}+c_5\cos^2{\theta}+c_6\cos{\theta}+c_7\right],
\end{equation}
where $c_4,\,c_5,\,c_6,\,c_7 \textrm{ and }d_4$ are functions of different dimensionless parameters. Again the complete expression of $ q^{(Re_{\!_E})}_{\,\!_S}$ can be found in the supplementary MATLAB file.

The streamfunction and surfactant distribution are obtained as:

\begin{equation}
\label{eq:ReE-sf}
\left.
\begin{split}
\Psi^{(Re_{\!_E})}_i & =\sum_{n=1}^{4}\left( A^{(Re_{\!_E})}_{{n}}{r}^{n+3}+B^{(Re_{\!_E})}_{{n}}{r}^{n+1} \right) Q_{{n}} \left( \eta\right) \\
\Psi^{(Re_{\!_E})}_e & ={r}^{2}U^{(Re_{\!_E})}_dQ_1(\eta)+\sum_{n=1}^{4}\left( C^{(Re_{\!_E})}_{{n}}{r}^{2-n}+D^{(Re_{\!_E})}_{{n}}{r}^{-n} \right) Q_{{n}} \left( \eta\right) \\
\end{split}
\right \}
\end{equation}

\begin{equation}\label{eq:surf-CaRe_E}
\Gamma^{(CaRe_{\!_E})}=\sum_{n=1}^{4} \Gamma^{(CaRe_{\!_E})}_nP_n(\eta)
\end{equation}
while the drop settling velocity takes the form:
\begin{subequations} \label{eq:Ud-ReE}
	\begin{equation}\label{eq:Ud-ReE-low}
	U^{(Re_{\!_E})}_d=-\frac{6}{5}\,
	{\frac 
		{M \left( 3\,R+3-S \right) 
			\left( R-S \right)  
		}{
		 \left( R+2\right) ^{2} \left( 2\,R+3 \right)  \left( 1+\lambda \right) \left( 2+3\,\lambda \right)  
		}
	}\times \underbrace{\left[\frac{(2+3\lambda)}
{(2+3\lambda)+\dfrac{k\beta}{1-\beta} }\right]^2}_\text{$k_1=$correction due to surfactant}
	\end{equation}
	In the High P\'eclet limit the same expression becomes:
	\begin{equation}\label{eq:Ud-ReE-high}
	U^{(Re_{\!_E})}_d=0
	\end{equation}
\end{subequations}
The physical significance of zero correction of $ O(Re_{\!_E}) $ drop velocity as appeared in the above equations is discussed  later in the results section.
Next we calculate the shape function of this order to investigate the surfactant-induced modifications on the deformation behaviour caused by charge convection. Mathematically,
\begin{equation}\label{eq:shapefunc_ReE}
f^{(CaRe_{\!_E})}=\sum_{n=2}^{4} L^{(CaRe_{\!_E})}_nP_n(\eta).
\end{equation}
The harmonics $L^{(CaRe_{\!_E})}_n $, being too cumbersome to present, are provided in the supplementary MATLAB file. 
\subsubsection{$O(Ca)$ solution}
\label{sss:O(Ca)_solution}
Solving the boundary conditions as outlined in section \ref{ss:perturbation}, we obtain the deformation induced modifications in the solution of electric problem as follows:
\begin{equation}
\label{eq:sol_pot-Ca}
\left.
\begin{split}
\varphi^{(Ca)}_i & =\sum_{n=1}^{3} r^{n}a^{(Ca)}_nP_n(\eta), \\
\varphi^{(Ca)}_e & = \sum_{n=1}^{3} r^{-n-1}b^{(Ca)}_{-n-1}P_n(\eta) , \\
\end{split}
\right \}
\end{equation}
where the non-zero harmonics are as follows:\\
\\
$a^{(Ca)}_1= -{\dfrac { 18\left( \,R-1 \right) }{5\, \left( R+2
		\right) ^{2}}}L^{(Ca)}_2,\, b^{(Ca)}_{-2}=\dfrac{6}{5}\,{\dfrac { \left( R-1 \right) ^{2}}{ \left( R+2
		\right) ^{2}}}L^{(Ca)}_{2} \textrm{ and }  \, 
	b^{(Ca)}_{-4}=\dfrac{9}{5}\,{\dfrac { \left( R-1 \right) }{R+2}}L^{(Ca)}_{2}.
 $
 \bigskip
\\Consequently the surface charge density is obtained as \bigskip
\begin{equation}\label{eq:sol-sc-Ca}
q^{(Ca)}_{\!_{\,S}}=\frac{6}{5}\,{\frac { \left( R-1 \right) \left( 15 \left( 
		\cos {\theta}  \right) ^{2}R+30 \left( \cos {\theta} \right) ^{2}-7R-3S-20 \right) \cos {\theta}}{ \left( R+2 \right) ^{2}}}L^{(Ca)}_{2}.
\end{equation}
Similarly the streamfunction and surfactant distribution turn out to be:

\begin{equation}
\label{eq:Ca-sf}
\left.
\begin{split}
\Psi^{(Ca)}_i & =\sum_{n=1}^{4}\left( A^{(Ca)}_{{n}}{r}^{n+3}+B^{(Ca)}_{{n}}{r}^{n+1} \right) Q_{{n}} \left( \eta\right) \\
\Psi^{(Ca)}_e & ={r}^{2}U^{(Ca)}_dQ_1(\eta)+\sum_{n=1}^{4}\left( C^{(Ca)}_{{n}}{r}^{2-n}+D^{(Ca)}_{{n}}{r}^{-n} \right) Q_{{n}} \left( \eta\right) \\
\end{split}
\right \}
\end{equation}

\begin{equation}\label{eq:surf-Ca2}
\Gamma^{(Ca^2)}=\sum_{n=1}^{4} \Gamma^{(Ca^2)}_nP_n(\eta)
\end{equation}
while the drop settling velocity takes the form:
\begin{subequations}
	\begin{equation}\label{eq:low-O_Ca_vel}
	U^{(Ca)}_d=\frac{1}{5}U^{(0)}_d(L^{(Ca)}_2c_6+Mc_7),	\end{equation}
	where $c_6=c_6(R,\beta,k,\lambda)$ and $c_7=c_7(R,\beta,k,\lambda,S)$ are complex functions of physical properties and are given in the supplementary MATLAB file.
	In the High P\'eclet limit the same expression becomes:
	\begin{equation}\label{eq:high-O_Ca_vel}
	U^{(Ca)}_d=\frac{1}{5}U^{(0)}_dL^{(Ca)}_2=
\frac{3}{20}\left(1-{\frac{1}{3(1+\lambda)}}\right){\frac { M\left( {R}^{2}+2\,R-4\,S+1 \right) }{{R}^{2}+4\,R+4}} 
	\end{equation}
\end{subequations}

We now determine the shape function of this order to investigate the surfactant-induced change in the higher order deformation behaviour. Mathematically,
\begin{equation}\label{eq:shapefunc_Ca}
f^{(Ca^2)}=L^{(Ca^2)}_0+\sum_{n=2}^{4} L^{(Ca^2)}_nP_n(\eta)
\end{equation}
Again the harmonics $L^{(Ca^2)}_n $ are provided in the supplementary MATLAB file. 
 Apart from these spherical harmonics a constant term ($ L^{(Ca^2)}_0 $) has to be included in the higher order deformation shape function to ensure the conserved volume of fluid. After applying the volume conservation constraint \citep{Mandal2016b}, we obtain 
 \begin{equation}\label{const-deform}
L^{(Ca^2)}_0=-\frac{1}{5}
\left(L^{(Ca)}_2\right)^2.
\end{equation}

 The surfactant concentration can be expresses as:
 \begin{equation}\label{eq:surf-Ca^2}
 \Gamma^{(Ca^2)}=\Gamma^{(Ca^2)}_0+\sum_{n=1}^{4} \Gamma^{(Ca^2)}_nP_n(\eta)
 \end{equation}
 The term $\Gamma^{(Ca^2)}_0$
 can be obtained from the mass constraint equation (\eqref{mass-constraint}). However in a motive to represent the surfactant distribution on a spherical surface, we employ a deformed-to-spherical surface transformation for $\Gamma$ through a projection operator as previously used by \citet{Vlahovska2005,Das2017a}:
 \begin{equation}
 \label{project}
 \Gamma_{\!_S}=\Gamma\frac{r_{\!_S}^2}{\mathbf{e}_r\cdot \mathbf{n}}
 \end{equation} where $\Gamma_{\!_S}$ is the projected concentration on the spherical surface. This gives\,
 
 \begin{equation}\label{eq:gamma_0_Ca2}
 \Gamma^{(Ca^2)}_0=-\frac{2}{5}\left(L^{(Ca)}_2\right)^2-\frac{2}{5}L^{(Ca)}_2\Gamma^{(Ca)}_2.
 \end{equation}
 
To check the validity of the present analytical calculations, we compared the above solutions with previous theoretical and experimental works for  different limiting conditions (please refer to Appendix~\ref{App-validation}).
  {\bigskip}

\section{Results and discussions}
In this section we show the effects of surfactant  on the settling behaviour of a viscous drop modulated by charge convection and shape deformation phenomena.
As a representative example we consider a leaky dielectric drop in another leaky dielectric surrounding fluid. Following the experiments of \cite{Mhatre2013} we adopt the physical  property values ($R=0.02, S=0.5638$ and $\lambda= 0.4359
$) corresponding to the situation when  a silicon oil drop is settling through a castor oil media which we denoted as system-A. Unless otherwise mentioned the property ratios of system-A are chosen along with $ k=1,\, M=5,\,Re_{\!_E}=0.2$ and $Ca=0.2 $. However in certain cases, to highlight the contrast in results caused by a different combination of electrohydrodynamic property ratios, we adopt another system with  ($ R=2.5,\,S=0.1  $ and $ \lambda=0.1 $) and denote as system-B.  Based on the practical consideration of different drop dimensions and applied electric field strengths, other  dimensionless parameters are calculated. 

In the present work, the influence of the surfactants on the interface tension appears through the elasticity parameter  $\beta$ and the property constant $k$. The uniform coating of surface active agents can reduce the interfacial tension significantly as compared to a clean surface via the relation $\gamma_{eq}=\gamma_c(1-\beta)$. A uniformly coated drop corresponds to the case $ Pe=0 $ $(\text{or }k=0) $ which physically signifies the fact that the surface diffusion is so dominant that the surfactant gradient becomes negligibly small. On the other hand $ \beta=0 $ corresponds to the condition when the surface tension is not at all sensitive to changes in local surfactant concentration. Here it is to be noted that  although we study the effect of the elasticity parmeter $ \beta $ by varying it upto 0.8 following \cite{Stone1990,Li1997}, the practicality of such high value of the parameter was questioned by many investigators \cite{Pawar1996,Eggleton1999} owing to assumed linear relationship in \eqref{eq:linear}. But under some instances, especially where the surfactant concentration is small, such values of $\beta$ can indeed closely predict the actual phenomena \cite{Li1997}. In what follows, we will highlight both of these parametric effects on flow and deformation characteristics of the drop. In the following discussions, we will highlight both of these parametric effects on flow and deformation characteristics of the drop. 

\subsection{Effects of surfactant on a spherical drop}
\label{subsec:Ud}
We first focus our attention towards the condition where the drop deformation is insignificant and the combination of  settling and EHD is affected by charge convection alone. It can be noted from \eqref{eq:Ud-ReE} that the charge convection induced effect on settling velocity vanishes at high convection limit ($ Pe \gg 1 $). In this limit the surface of the drop becomes completely immobilized by the surfactant molecules and the surface velocity tends to zero, rendering the charge convection effect to become negligible too.

\begin{figure}[!htbp]
	\centering
	\begin{subfigure}[!htbp]{0.45\textwidth}
		\centering
		\includegraphics[width=\textwidth]{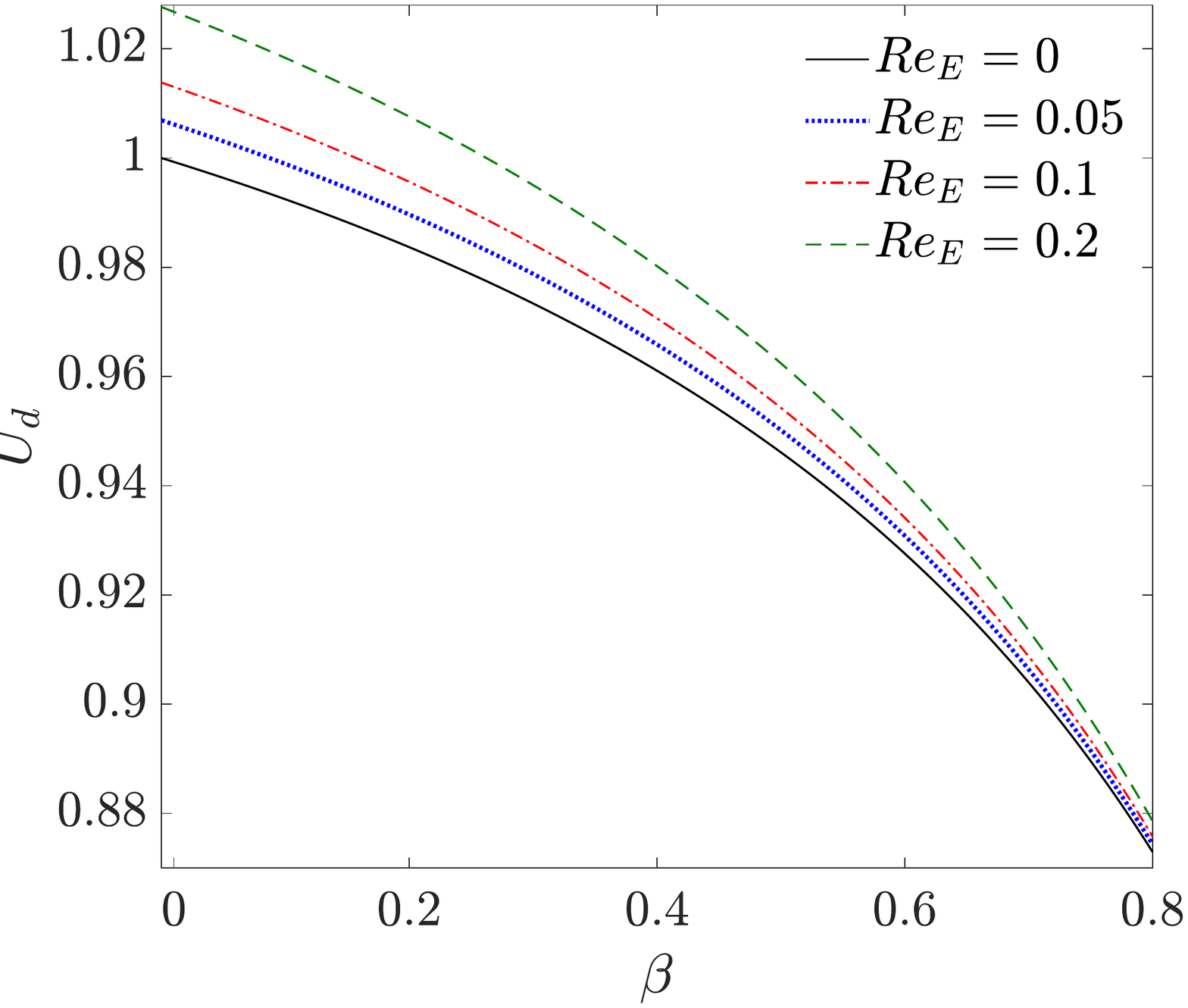}
		\caption{}
		\label{fig:Ud-ReE-A}
	\end{subfigure}
	\quad 
	\begin{subfigure}[!htbp]{0.45\textwidth}
		\centering
		\includegraphics[width=\textwidth]{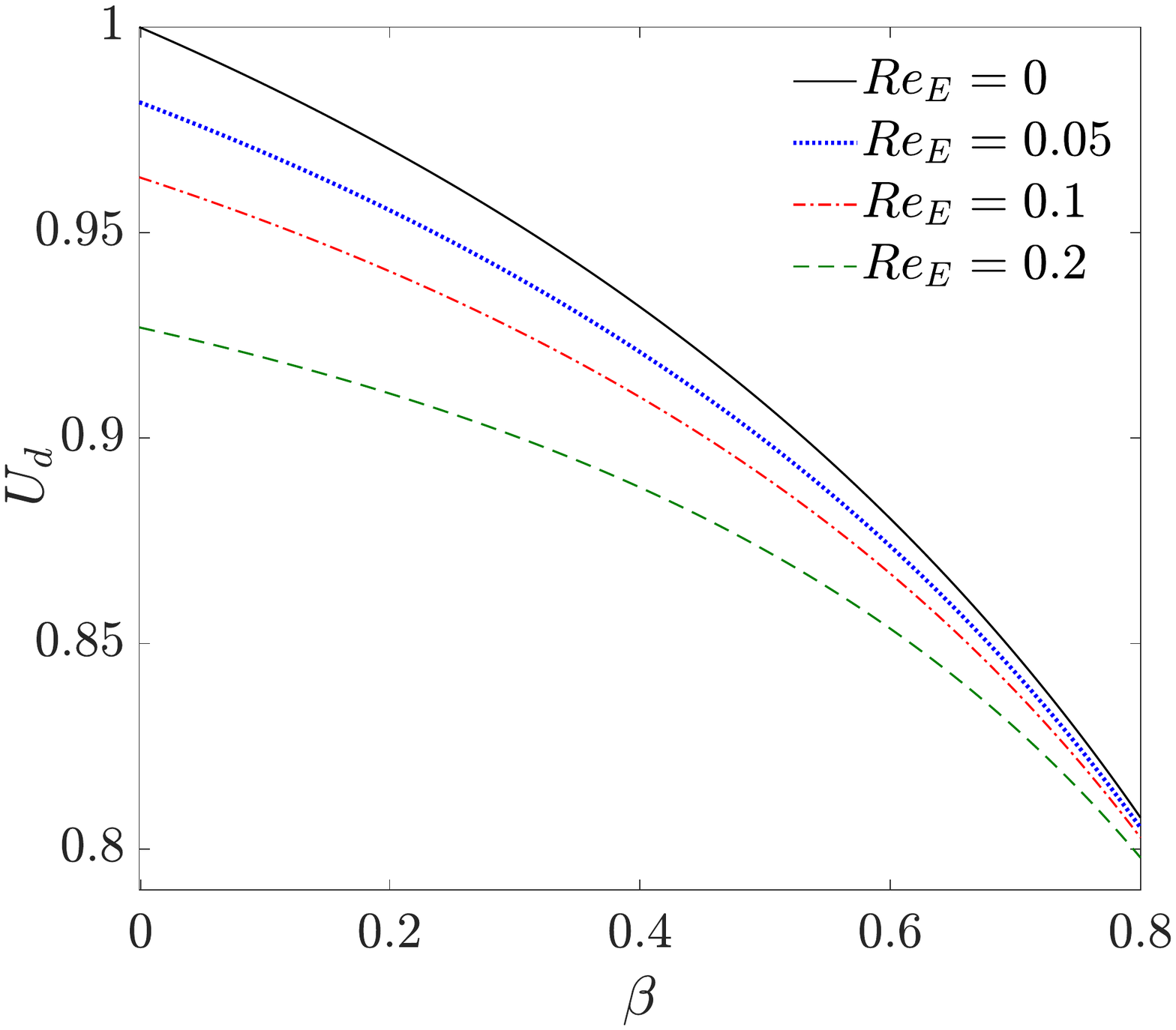}
		\caption{}
		\label{fig:Ud-ReE-B}
	\end{subfigure}
	\vspace*{0em}
	\caption{Settling velocity $ U_d $ vs. $\beta$ for different electric Reynolds number ($  Re_{\!_E}$). Subplot (a) is for system-A and (b) for system-B. Other parameters are $k=1$ and $ M=5 $.}
	\label{fig:Ud_ReE_beta}. 
\end{figure}

In figure\,\,\ref{fig:Ud_ReE_beta}  the variation of drop settling velocity $ U_d=U^{(0)}_d+Re_{\!_E} U^{(Re_{\!_E})}_d$ with the elasticity parameter $ \beta $ has been shown for different values of $ Re_{\!_E} $. It is observed that $ U_d $ decreases continuously with increasing $ \beta $ for both system-A and B, while the effect of $ Re_{\!_E} $ is just the opposite  for the two cases. It is also interesting to find that charge convection modulated velocity deviation becomes extremely small at higher values of $ \beta $. The corresponding behaviour can be  understood by noting the expression of $U^{(Re_{\!_E})}_d$ in 
\eqref{eq:Ud-ReE-low}.  The discriminating factor $ \mathcal{F}=\left( 3\,R+3-S \right) 
\left( R-S \right) $, which determines the sign of change in drop settling velocity modification by charge convection ($ \mathcal{F}<0  $ for system-A and $ \mathcal{F}>0 $ for system-B), remains unchanged even if the surfactant effects are present. However, the presence of surfactants can alter the amount of correction ($ k_1 $) to $ U^{(Re_E)}_d $ determined by a coupling of viscosity ratio $ \lambda $ with surfactant parameters $ \beta $ and $ k $.  For a mathematical  justification of the above behaviour, we revisit the correction factor $ k_1 $ appeared in \eqref{eq:Ud-ReE}.  Keeping in view of the practical range of various parameters, it is found that $ k_1 \leq1 $ condition is satisfied always and hence the magnitude of $ U^{(Re_{\!_E})}_d $ decreases due to surfactants. It is identified from the present calculation that  the electric force $ \mathbf{F}^{E(Re_{\!_E})} $ turns out to be zero and thus the hydrodynamic force $ \mathbf{F}^{H(Re_{\!_E})} $ is the sole responsible for the velocity correction due to charge convection.

\begin{figure}[!htbp]
	\centering
	\begin{subfigure}[!htbp]{0.4\textwidth}
		\centering
		\includegraphics[width=\textwidth]{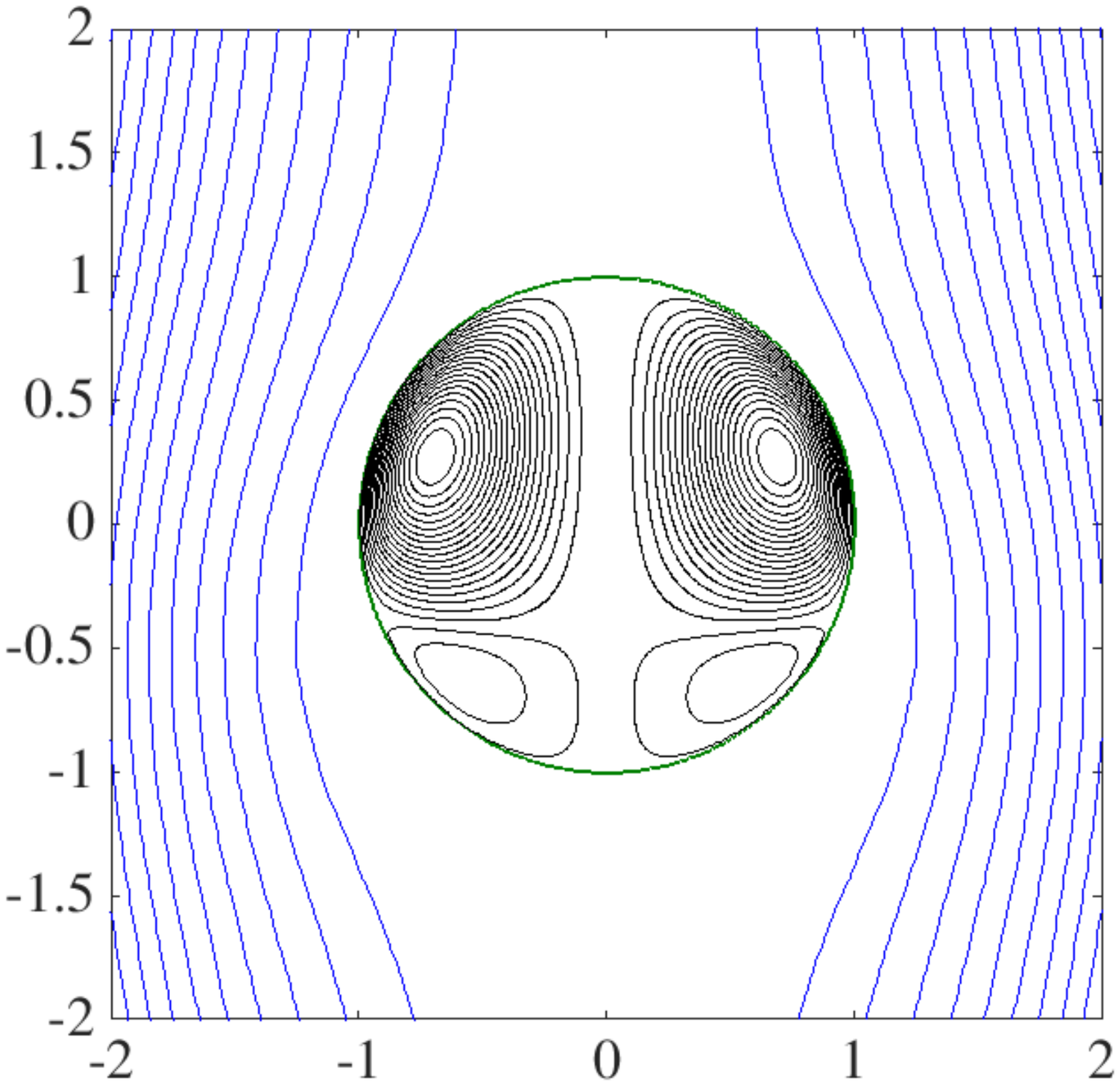}		\caption{}
		\label{fig:contour-clean}
	\end{subfigure}
	\quad %Horizontal spacing commands.enskip, quad, qquad leave a horizontal space of respectively half an em, one em and two ems. The "em" is a font depending length, frequently as wide as a capital M in the current font.
	\begin{subfigure}[!htbp]{0.4\textwidth}
		\centering
		\includegraphics[width=\textwidth]{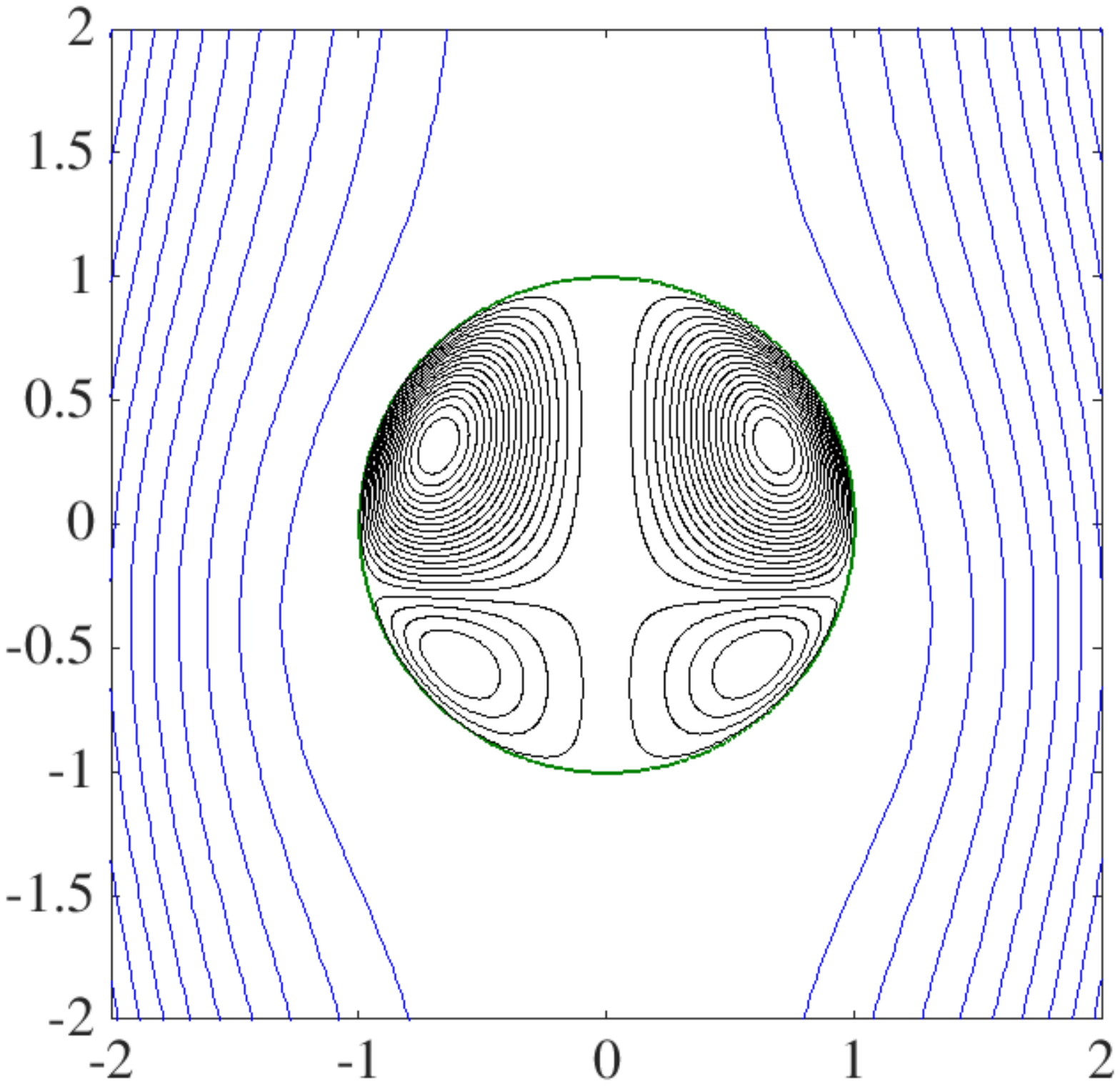}
		\caption{}
		\label{fig:contour-surf}
	\end{subfigure}
\\[-12ex] 
\begin{subfigure}[!htbp]{0.45\textwidth}
	\centering
	\includegraphics[width=\textwidth]{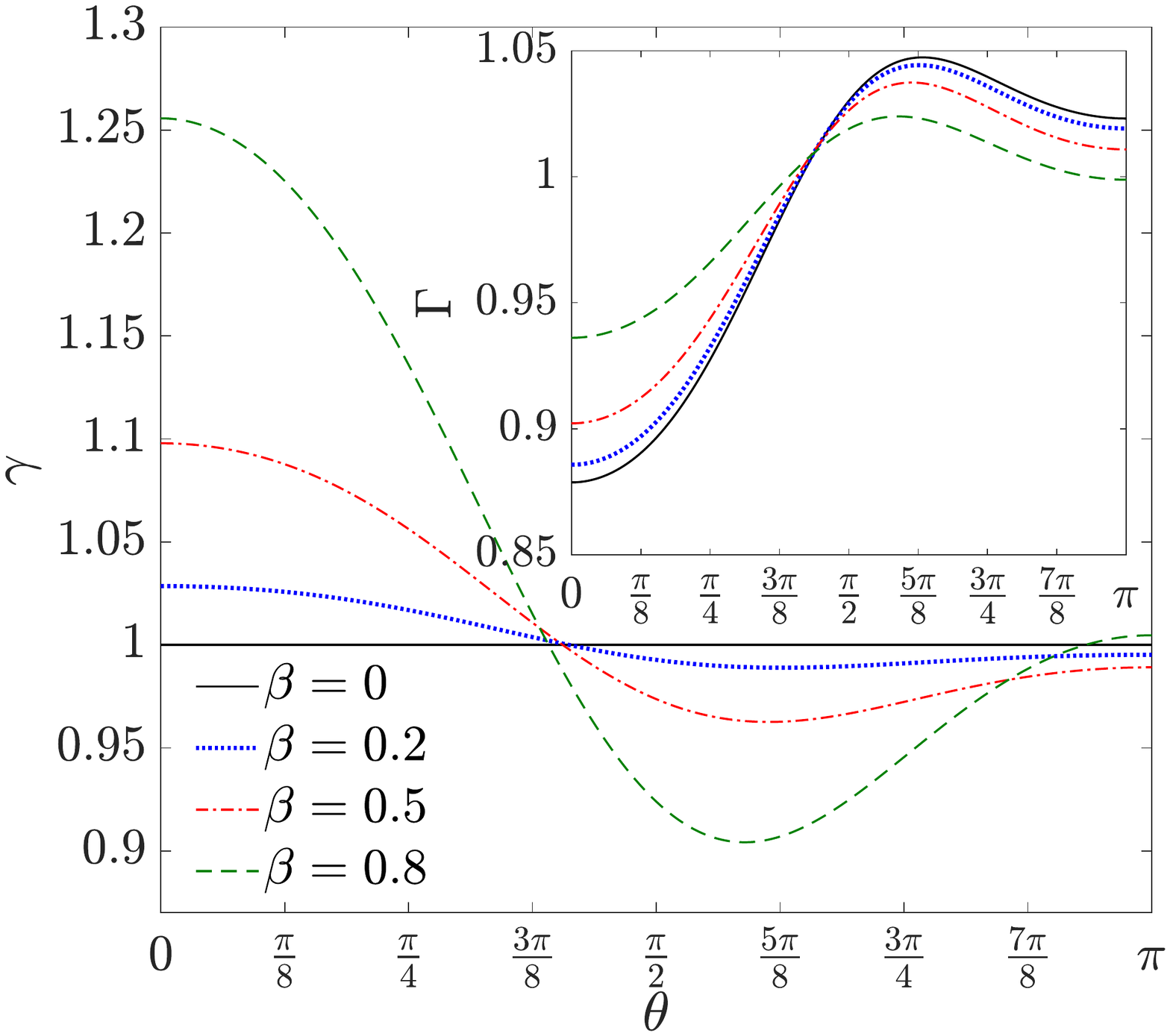}
	\caption{}
	\label{fig:surf_beta}
\end{subfigure}
	\\[-12ex] %Add something like this just after the second subfigure reduces the distance between the first and second rows.
	\begin{subfigure}[!htbp]{0.45\textwidth}
		\centering
		\includegraphics[width=\textwidth]{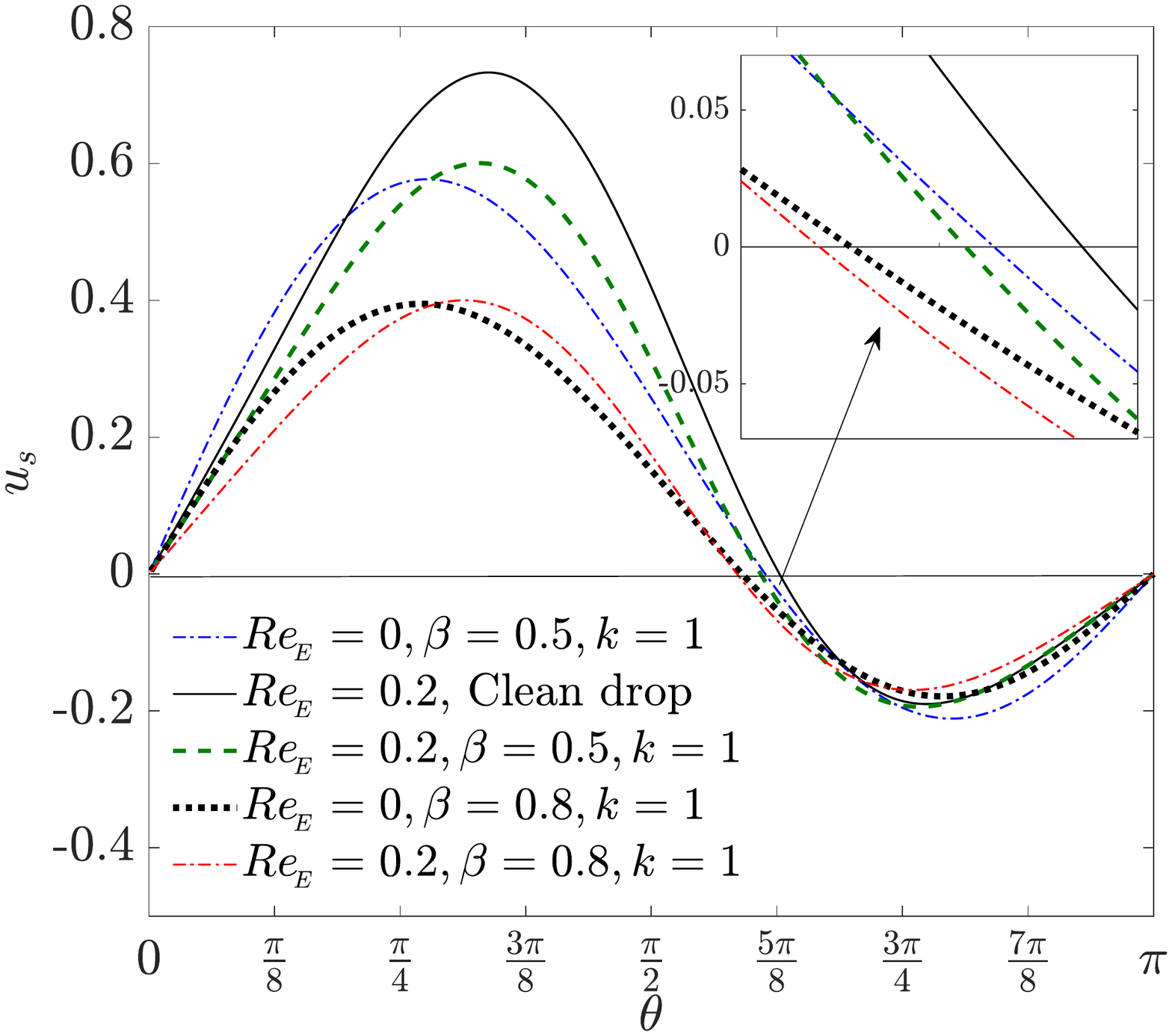}
		\caption{}
		\label{fig:us_vary_Re_beta_0p8_inset}	
	\end{subfigure}
%\\[-12ex]
	\begin{subfigure}[!htbp]{0.45\textwidth}
		\centering
		\includegraphics[width=\textwidth]{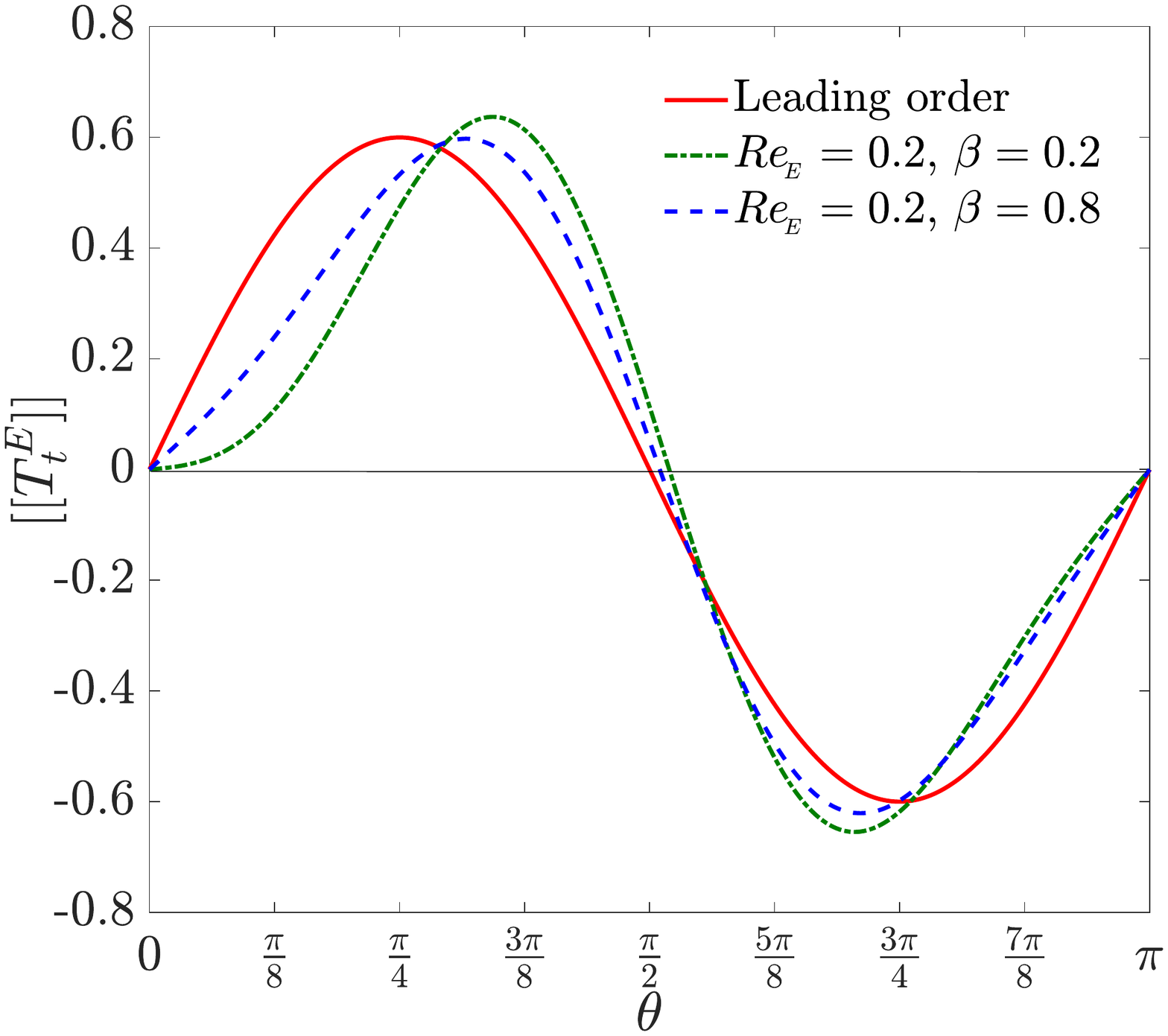}
		\caption{}
		\label{fig:tangential_vary_Re_beta}
	\end{subfigure}
%\\[-12ex] 
%	\begin{subfigure}[!htbp]{0.45\textwidth}
%	\centering
%	\includegraphics[width=\textwidth]{surf_vary_beta_order_RE}
%	\caption{}
%	\label{fig:surf_beta}
%\end{subfigure}
	\caption{Streamline pattern for (a) clean drop and (b) for surfactant coated case of the same drop. Subplot (c) shows surface tension distribution ($ \gamma $) for  different values of $ \beta $; (d) shows the interfacial velocity distribution  different values of $ \beta , k \, \text{and} \, Re_{\!_E}$; (e) shows the tangential electric stress jump for different values of $ \beta $.}
	\label{fig:effect_beta_ReE_Us_Tt}. 
\end{figure}

 To obtain a comprehensive physical picture, we plot the streamline pattern, surface velocity and tangential component of electric stress jump in figure\,\,\ref{fig:effect_beta_ReE_Us_Tt}. From figure\,\,\ref{fig:contour-clean} it can be observed that an asymmetric flow pattern about the equator builds up due to redistribution of charges by charge convection phenomenon. As a consequence circulation cells, with flow direction from the poles to the stagnation points, set in. This being the picture for a surfactant-free drop, presence of surface-active agents on the interface brings Marangoni stress in competition to the electric stress.  In figure\,\,\ref{fig:contour-surf} the change in flow pattern due to surfactant is highlighted.
To identify the behaviour of the Marangoni stress and determine its traction direction we plot the surface tension and surface concentration distribution for different values of $ \beta $ in figure\,\,\ref{fig:surf_beta}. It is found that increase in $ \beta $ causes a significant variation of surface tension from its equilibrium value and the driving force for the Marangoni flow, i.e. the surface tension gradient $|\gamma_{max}-\gamma_{min}|$, increases. 
From the inset of the same figure we find that the surface concentration tends to reach the equilibrium concentration as $ \beta $ increases. This apparently contradicting behaviour of surface concentration  with respect to surface tension can be explained with a  due consideration of the underlying physical mechanisms.  With an increase in $ \beta $ the surface tension gradients become strong enough to restrict convective transport of surfactant molecules towards the stagnation points and gradient in surface concentration $|\Gamma_{max}-\Gamma_{min}|$ gets reduced. Similar qualitative observations were also made in case of a viscous drop in linear shear by \citet{Li1997}. 
Now according to the traction directions determined by the surface tension gradient, Marangoni stress tries to  trigger a fluid flow towards polar region, thus opposing the electrohydrodynamically induced flow. The change in flow pattern in figure\,\,\ref{fig:contour-surf} can be justified by the shift in stagnation points which can be visualized from the surface velocity plots in figure\,\,\ref{fig:us_vary_Re_beta_0p8_inset}. As $ \beta $ increases, the stagnation points shift more towards the front pole ($ \theta=0 $). Similar trend in the shift of sign reversal point of the tangential electric stress jump $T^E_t$, is also found in  figure\,\,\ref{fig:tangential_vary_Re_beta}. As a consequence magnitude of the vortices near the front pole increases. This creates alteration in internal flow structure and the resulting hydrodynamic drag force experienced by the drop.

In some of the representative cases discussed, modulations created by surfactant parameter variations are not visually distinguishable. However, a different combination of  parameters (e.g.\ $ R=0.01,\,S=1.7  $, $ \lambda=0.03  $ and $ M=2.7 $) following \cite{Xu2006} (please refer to Appendix~\ref{sec:App-1}) reveals the variations in a more clear and distinguishable manner as shown in figure\,\,\ref{fig:Xu-Homsy-Streamilne-Stress}. Since in that case, discriminating factor $ \mathcal{F} $ remains positive, resultant effect of $ \beta $ on the charge convection induced velocity increase gives the same trend as of figure\,\,\ref{fig:Ud-ReE-A} (not shown here for brevity). It is interesting to observe that such a parameter choice gives rise to additional two pairs secondary circulation cells near the poles in the absence of surfactant molecules (please refer to figure\,\,\ref{fig:Xu_contour}). Now with an increase in the elasticity parameter $\beta$ makes the secondary rolls become diminishingly small (shown in figure\,\,\ref{fig:Xu_contour_surf}). The corresponding prominent modulations in $ T^E_t $ are also depicted in figure\,\,\ref{fig:Homsy_TE_t_vary_beta}. The surfactant induced modifications in the internal circulation rolls, as discussed above, can severely alter the mixing characteristics inside the drop. 

\begin{figure}[!htbp]
	\centering
	\begin{subfigure}[!htbp]{0.52\textwidth}
		\centering
		\includegraphics[width=\textwidth]{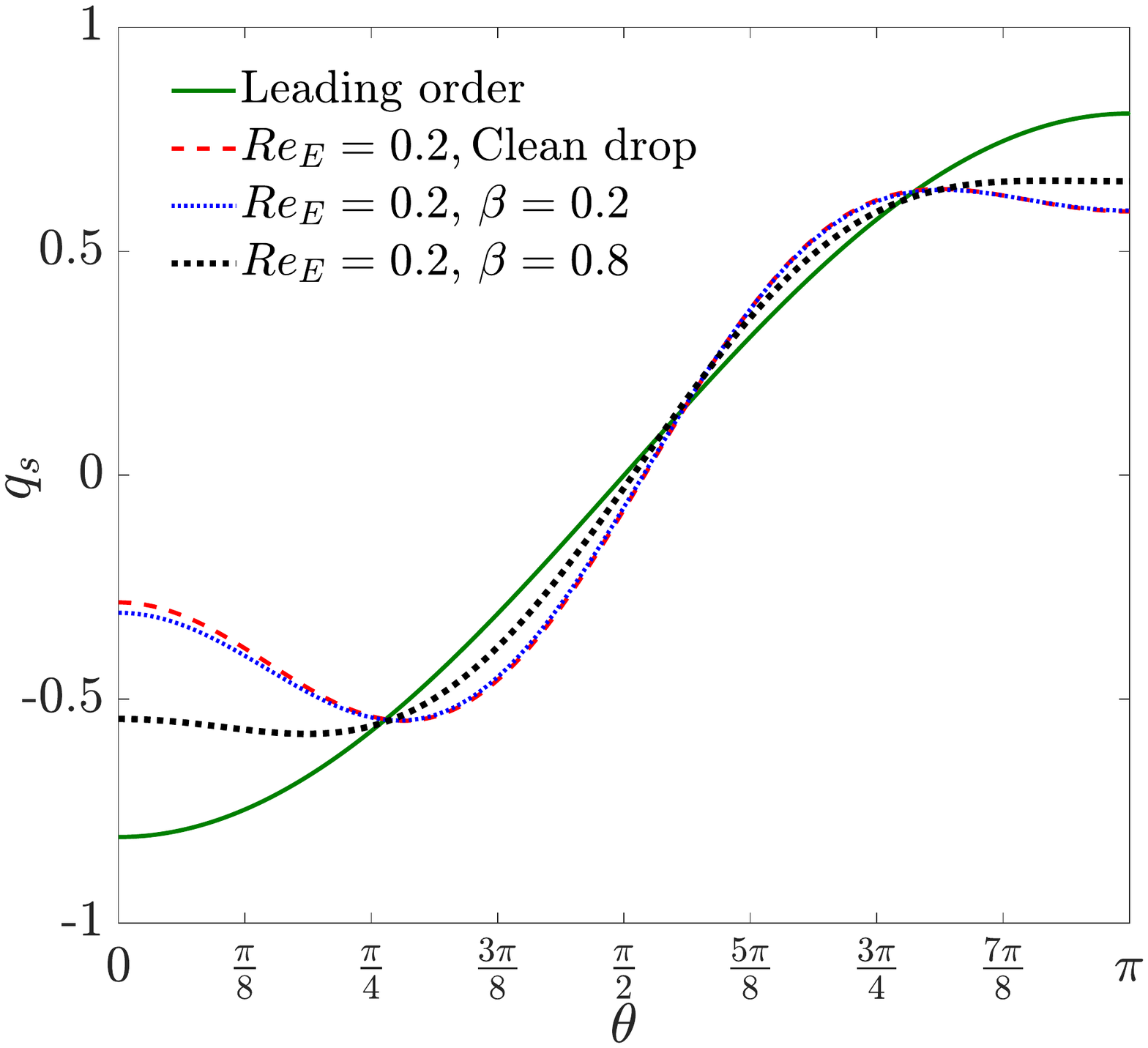}
		\caption*{}
		\label{fig:charge_ReE_beta}
	\end{subfigure}
	\vspace*{0em}
	\caption{ Surface charge distribution for leading-order as well as including charge convection effect ($ q_{\,\!_S}=q^{(0)}_{\,\!_S}+ Re_{\!_E}q^{( Re_{\!_E})}_{\,\!_S}$) for various values of $\beta$.}
%	 (b) Leading-order surface tension variation with surfactant concentration in the inset.
	\label{fig:charge_ReE_beta_AB}. 
\end{figure}

\begin{figure}[!htbp]
  \centering
  \begin{subfigure}[!htbp]{0.6\textwidth}
    \centering
    \includegraphics[trim=0 0 0 5pt,clip,width=\textwidth]{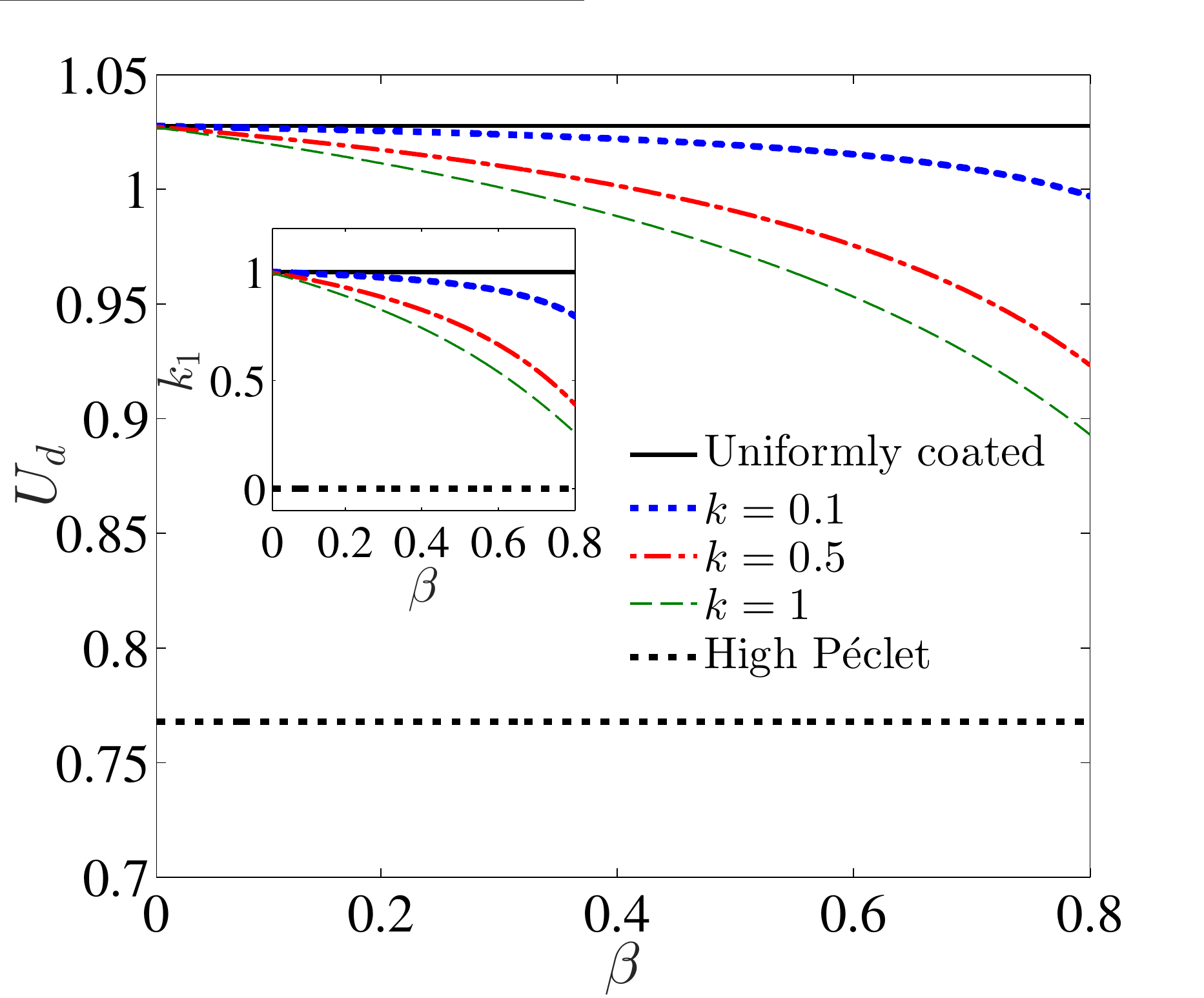}
    \vspace{70pt}
    \caption{}
    \label{fig:ud_vs_beta_vary_k_Re_0p2}		
  \end{subfigure}
  \\[-8ex]
  \begin{subfigure}[!htbp]{0.45\textwidth}
    \centering
    \includegraphics[width=\textwidth]{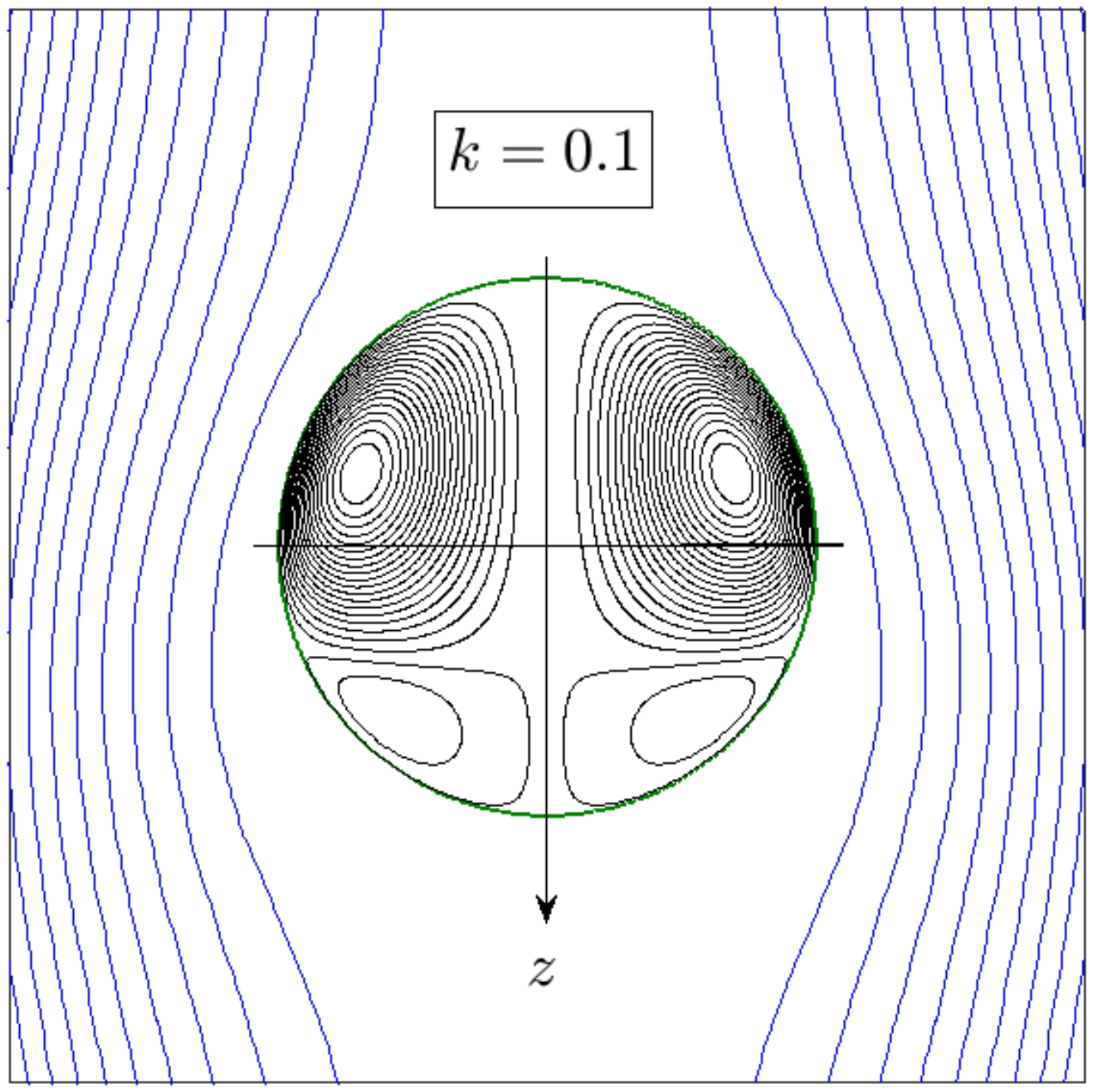}
    \caption{$ $}
    \label{fig:contour_low_k_high_beta}		
  \end{subfigure}
  \begin{subfigure}[!htbp]{0.45\textwidth}
    \centering
    \includegraphics[width=\textwidth]{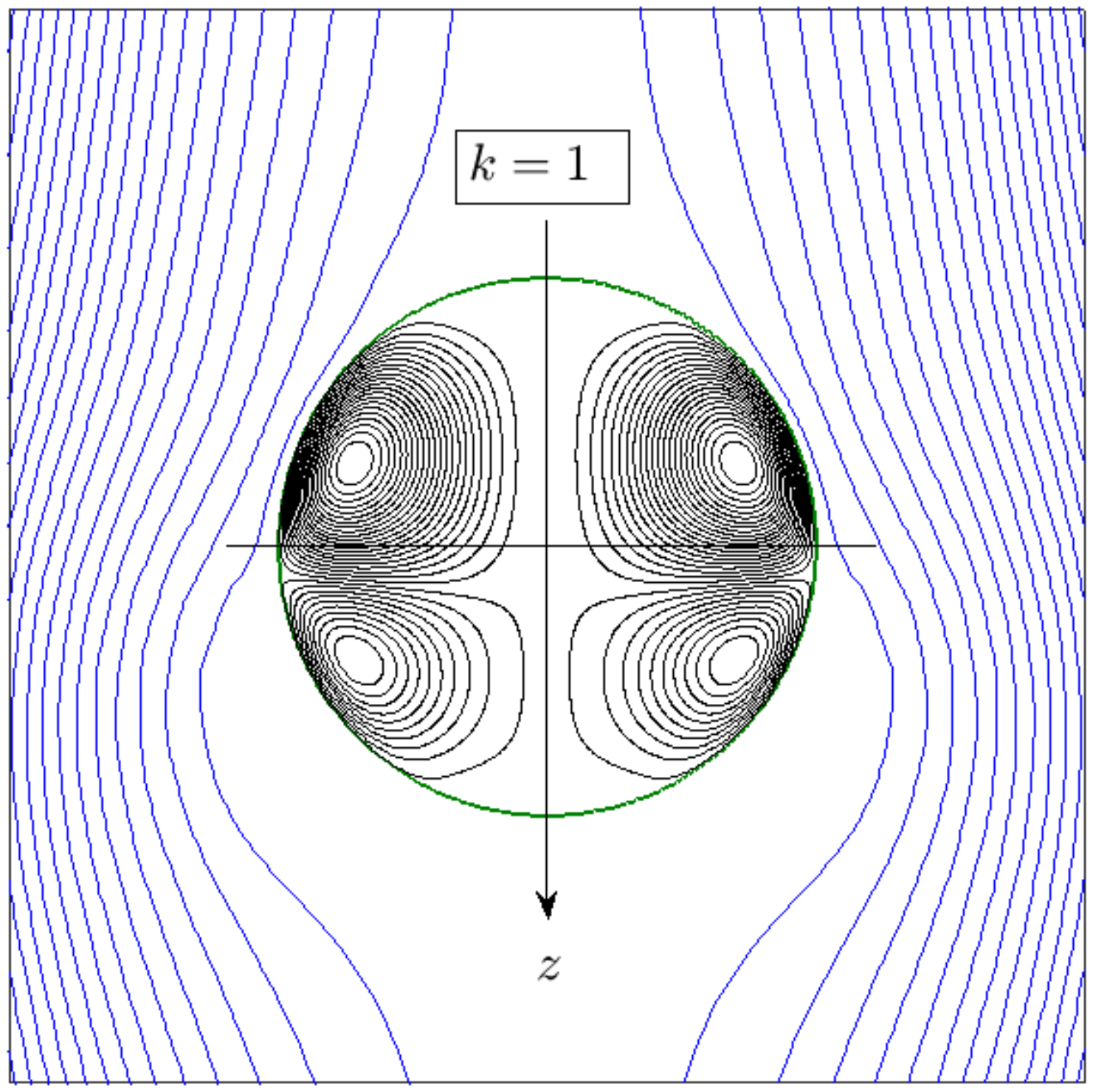}
    \caption{}
    \label{fig:contour_high_k_high_beta}		
  \end{subfigure}
  \caption{(a) Variation of drop settling velocity $ U_d $ with elasticity parameter $ \beta $ for various values of property ratio $ k $. Subplots (b) and (c) are the streamline patterns for $ \beta=0.8 $ and two different $ k $.}
  \label{fig:contour_vary_k}. 
\end{figure}

The intricate behaviour of the flow not only affects the drop velocity but also causes significant change in charge distribution as described in figure~\ref{fig:charge_ReE_beta_AB}. 
 For a surfactant-free drop the smaller rolls near the front pole possess more strength of flow as compared to the primary rolls adjacent to the rear pole. Thus the charges are more dominantly swept away from front pole towards the stagnation point as compared to that is achieved by the primary rolls adjacent to the rear pole. Such behaviour qualitatively agrees with the work of \cite{Mandal2016a} for a neutrally buoyant drop motion driven by dielectrophoresis. The presence of  surfactants weakens the strength of smaller rolls near the poles, thus diminishing the process of convection of charges towards the stagnation points. 
%\vspace*{-4em}
Figure \ref{fig:ud_vs_beta_vary_k_Re_0p2} depicts the effect of the property ratio $ k $ on the drop settling velocity.	In the extreme limit of $ k=0 $ the drop is uniformly coated with surfactant. Consequently the correction term in leading-order drop velocity becomes zero while the factor $ k_1 $ in $ O(Re_{\!_E}) $ drop velocity becomes 1. Thus the drop velocity becomes same as that of a surfactant-free drop given by
\begin{equation}\label{eq:Ud-leading-low-Pe-lambda}
\lim_{k \to 0}	U_d=1-\frac{6}{5}\,
{\frac 
	{M \left( 3\,R+3-S \right) 
		\left( R-S \right)  
	}{
		\left( R+2\right) ^{2} \left( 2\,R+3 \right)  \left( 1+\lambda \right) \left( 2+3\,\lambda \right)  
	}
}
\end{equation}
  As seen from the figure in the low P\'eclet number regime, with an increase in $ k $ the drop velocity decreases as compared to the surfactant-free drop while in the high P\'eclet regime it becomes insensible to the parameter $ \beta $  and finally reaches the non-diffusing limit set by \eqref{eq:Ud-leading-high}. For the low P\'eclet limit when surface advection of surfactant molecules is relatively weak, redistribution of charges for non-zero $ Re_{\!_E} $  can significantly alter the drop velocity. It can be inferred from the correction term $ k_1 $ in \eqref{eq:Ud-ReE-low} that increase in $ k $ will always reduce the magnitude of the charge convection induced correction to drop velocity. The functional nature of the correction factor $ k_1 $ is shown in the inset of figure\,\,\ref{fig:ud_vs_beta_vary_k_Re_0p2}. The primary effect of increasing $ k $ is to increase the  surface P\'eclet number and thus it should have increased the charge convection effect due to enhanced EHD induced convection of fluid. But the secondary competing effect of increased interfacial tension gradient due to increased $ k $, dominates in this case, and the opposing Marangoni flow intensifies. Such behaviour in the low P\'eclet limit is similar to that caused by the elasticity parameter $ \beta $. It can be verified  from the streamline structures presented in 	figure\,\,\ref{fig:contour_low_k_high_beta}		and		\ref{fig:contour_high_k_high_beta}. Similar to the effect of increasing $ \beta $,  size of the secondary vortices increases with increasing $ k $. Another important observation from figure\,\,\ref{fig:ud_vs_beta_vary_k_Re_0p2} is that as $ \beta $ increases, the retardation effect initiated by increasing $ k $, gets intensified. This can be attributed to a combined additive influence of $ k $ and $ \beta $ on the enhancement of Marangoni flow opposite to the EHD flow.  

The  effect of viscosity ratio on the surfactant induced modification in drop velocity, in absence of charge convection, has been discussed in earlier studies \citep{Teigen2010,Mandal2016a}. Following a similar trend the leading-order correction to drop velocity brought by surfactant, in both the low and high P\'eclet limits become maximum  when the drop viscosity is negligible with respect to the outer medium ($ \mu_i \ll \mu_e $). Such a condition closely reduces to the case of a bubble in an otherwise viscous medium. The limit is obtained as a correction to the Hadamard-Rybczynski velocity as:
\begin{subequations}
	\begin{equation}\label{eq:particle-Ud-low-leading}
	\lim_{\lambda \to 0}	U^{(0)}_d=1-
	\underbrace{\frac{1}{3}\,
          {\frac 
                  {k\,\beta}
			{ 
                          \left(2(1-\beta)+k\beta\right)}}
	}
	_\text{correction due to surfactant}
	\end{equation}
	while in the High P\'eclet limit the same expression becomes:
	\begin{equation}\label{eq:particle-Ud-high-leading}
	\lim_{\lambda \to 0}	U^{(0)}_d=1-\underbrace{\frac{1}{3}}_{\substack{\text{correction due} \\ \text{to surfactant}}}
	\end{equation}
\end{subequations}
In a similar manner the $ O(Re_{\!_E}) $ drop velocity ($ U^{(Re_{\!_E})}_d $) takes a maximum value under such a  condition given by:
\begin{equation}\label{bubble_Ud_ReE}
\lim_{\lambda \to 0} U^{(Re_{\!_E})}_d = -\frac{3}{5}{\frac { \left( R-S \right)  \left( 3\,R-S+3 \right) M}{ \left( 
		2\,R+3 \right)  \left( R+2 \right) ^{2}}}\times  \underbrace{\frac {4 \left( 1-\beta \right) ^{2}}{ \left( 2\left( 1-\beta \right)+  k
		\beta \right) ^{2}}}_\text{$ k_1= $ correction due to surfactant}.
\end{equation} 
It is to be noted here that the correction term due to surfactant $ k_1 $ also becomes maximum in that limit.
On the other hand when the drop becomes highly viscous  ($ \mu_i \gg \mu_e $) the surface velocity vanishes resembling the case of a solid sphere. In this case the internal circulations of the drop are already retarded too much due to viscous effects that the additional Marangoni effect has hardly any additional retardation effect to bring in. Thus the leading-order surfactant correction becomes zero while the $ O(Re_{\!_E}) $ drop velocity vanishes altogether.

The Mason number ($ M $) which depicts the relative importance of electric stress to that of hydrodynamic stress,  enhances the effect of charge convection on the drop velocity and the corresponding alterations by surfactant. This can be understood from the $ O(Re_{\!_E}) $ velocity perturbation in \eqref{eq:Ud-ReE}. 
\subsection{Effects of surfactant on a deformable drop}
\label{subsec:Deformation}
In this section we investigate how the deformability of droplet, affected primarily by surfactant parameters $ \beta$ and $k $, can influence the drop settling velocity and resulting drop shape.
 Here we consider only the $ O(Ca) $ correction to distinguish from the $ O(Re_{\!_E}) $ modifications as discussed in the preceding section. 
\begin{figure}[!htbp]
	\centering
	\includegraphics[width=0.65\textwidth]{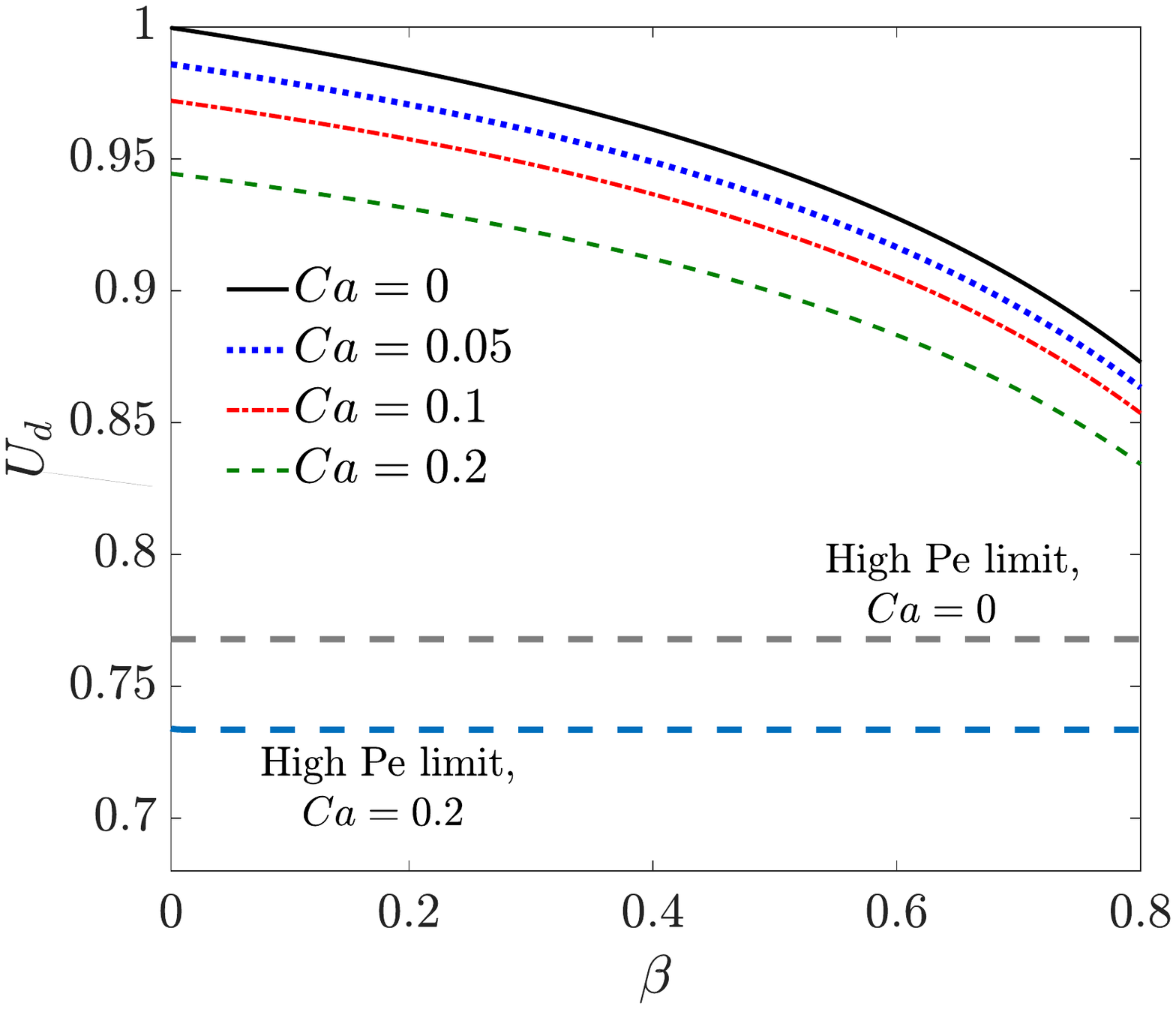}
	\vspace{-120pt} 
	\caption{ Variation of velocity $ U_d $ with elasticity parameter ($ \beta $) for different values of the capillary number ($ Ca $) in the low P\'eclet limit with $ k=1 $. The high P\'eclet limits of velocity are also shown for two different $ Ca $.}
	\label{fig:Ud_vary_beta-Ca-inset}
\end{figure}
During drop deformation the transport of surfactants on the interface, driven by the internal flow field and Marangoni induced surface stresses, leads to a phenomenon called tip stretching. This refers to the mechanism in which accumulation of surfactant molecules, favored by the flow direction, causes  the local surface tension to reduce and thus making such localized areas on the drop surface more vulnerable to deformation. Accommodation of such effects in the present mathematical model comes through the second term  in the right hand side of the normal stress balance equation (\eqref{eq:n_stress_bal}). The intricate interplay among these competing physical mechanisms dictates the resulting modifications brought in by the presence of surfactants.

Figure\,\,\ref{fig:Ud_vary_beta-Ca-inset} describes the effect of the elasticity parameter ($ \beta $) on  drop settling velocity with increasing order of deformability. For a specific value of capillary number ($Ca$), increase in $ \beta $ causes the drop to deform more in a oblate spheroid shape. This represents higher and higher flow obstruction with elevated hydrodynamic drag and a corresponding decrease in drop velocity results. The figure also depicts that in the high $ Pe $ regime the drop velocity becomes insensitive to the parameter $ \beta $ but the velocity gets reduced significantly as compared to the low $ Pe $ cases. 

\begin{figure}[!htbp]
	\centering
	\begin{subfigure}[!htbp]{0.45\textwidth}
		\centering
		\includegraphics[width=\textwidth]{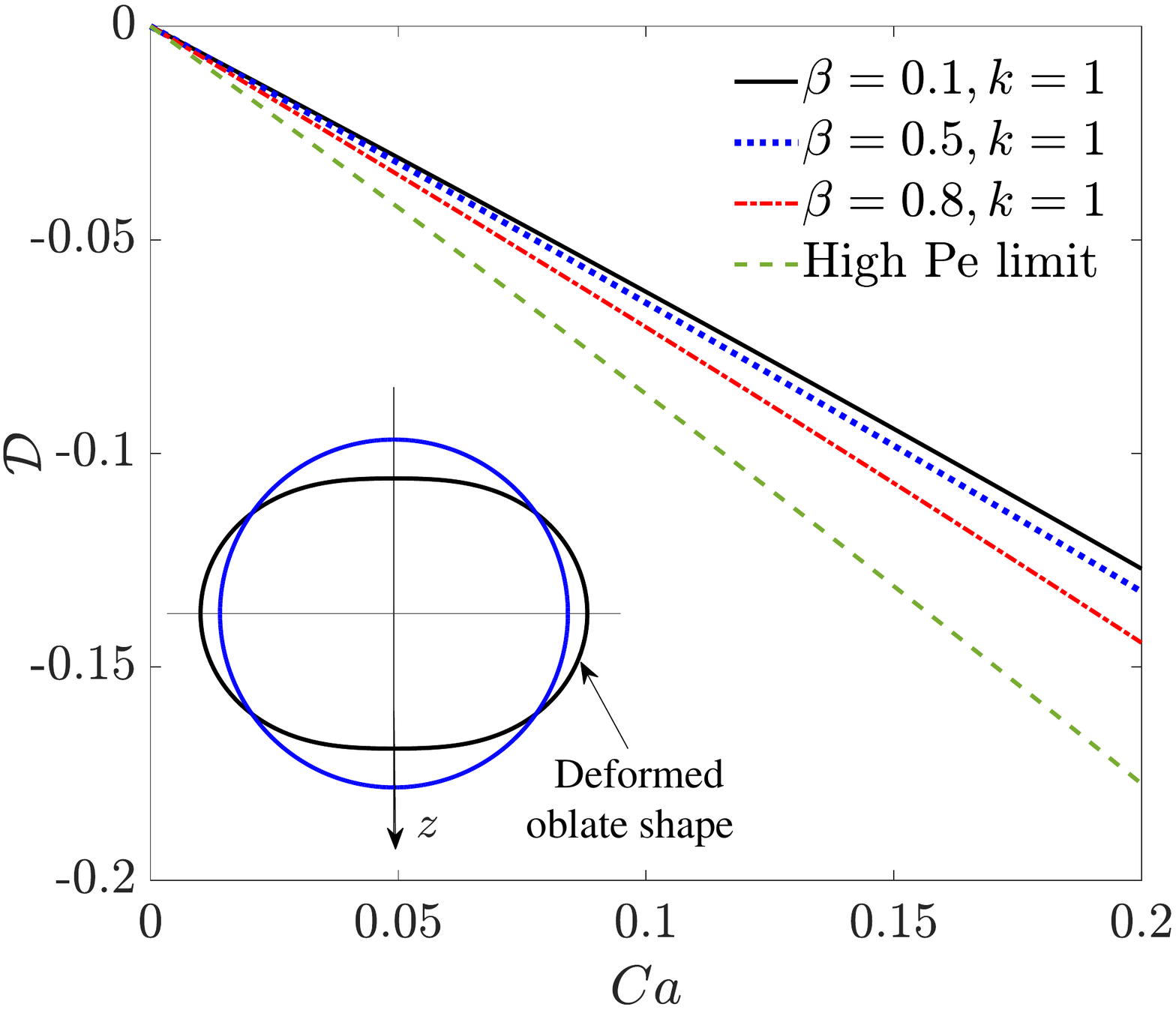}
		\caption{}
		\label{fig:D-vs-Ca-vary-beta}
	\end{subfigure}
\quad
	\begin{subfigure}[!htbp]{0.45\textwidth}
						\vspace*{5.0em}
		\centering
		\includegraphics[width=\textwidth]{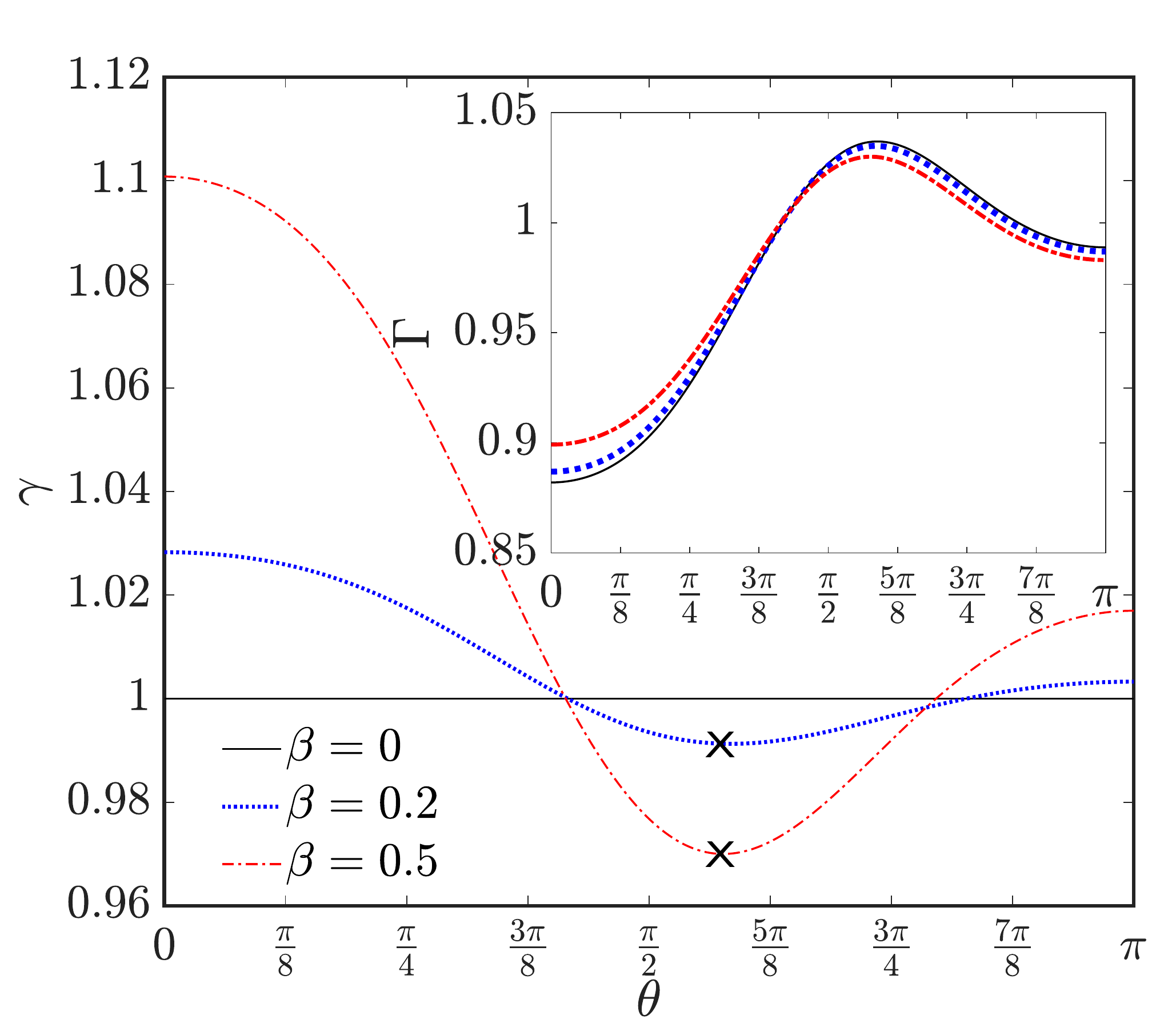}
%			\vspace{50pt} 
							\vspace*{4em}
		\caption{}
		\label{fig:tension-beta-Ca-high}	
	\end{subfigure}
	\vspace*{15pt}
	\caption{Subplot (a): Leading-order deformation parameter ($ \mathcal{D} $) vs. capillary number ($Ca$) for different values of elasticity parameter $\beta$. The high P\'eclet limit of deformation is also presented. In the inset the corresponding deformed oblate shape is shown for a representative case of high P\'eclet limit ; Subplot (b): Distribution of surface tension for different values of elasticity parameter $ \beta$. In the inset  we show the corresponding concentration variation.}
	\label{fig:D-tension-beta-Ca}.
\end{figure}

The behavior of  $ U_d $  in the high $ Pe $ limit can be substantiated by the enhanced oblate deformation of the drop (shown in figure\,\,\ref{fig:D-vs-Ca-vary-beta}). In figure\,\,\ref{fig:D-vs-Ca-vary-beta} the variation of higher order deformation parameter ($ \mathcal{D} $) with capillary number ($ Ca $) has been shown for different $ \beta $. 
Accumulation of surfactant molecules from the poles towards the stagnation points becomes more dominant in this case as compared the opposing effect of Marangoni flow. As a consequence the surface tension is lowest adjacent to the stagnation points (denoted by marker `${\times}$' in
figure\,\,\ref{fig:tension-beta-Ca-high}). In order to balance the local normal stresses generated due to electrohydrodynamics, the drop surface becomes more prone to deformation in these regions. This is reflected in figure\,\,\ref{fig:D-vs-Ca-vary-beta}.  It is also observed from the same figure that the surfactant effect on deformation gets enhanced as the capillary number ($Ca$) increases. This is due to the fact that increase in $Ca$ indicates a corresponding increase in surface P\'eclet number ($Pe=k\,Ca$) which pushed more surfactant molecules to these zones, causing the interfacial tension to reduce further. Now for system-B the drop takes a prolate spheroid shape. Since in this case the surfactant molecules are swept towards polar regions by the dominant action of electrohydrodynamic stress, the prolate behaviour is intensified. Thus the drag experienced by the drop is further reduced and finally it settles faster than the surfactant-free case (shown in figure\,\,\ref{fig:prolate-deform}). 

\begin{figure}[!htbp]
		\centering
%		\vspace*{-5em}
	\begin{subfigure}[!htbp]{0.435\textwidth}
%				\vspace*{0em}
		\centering
		\includegraphics[width=\textwidth]{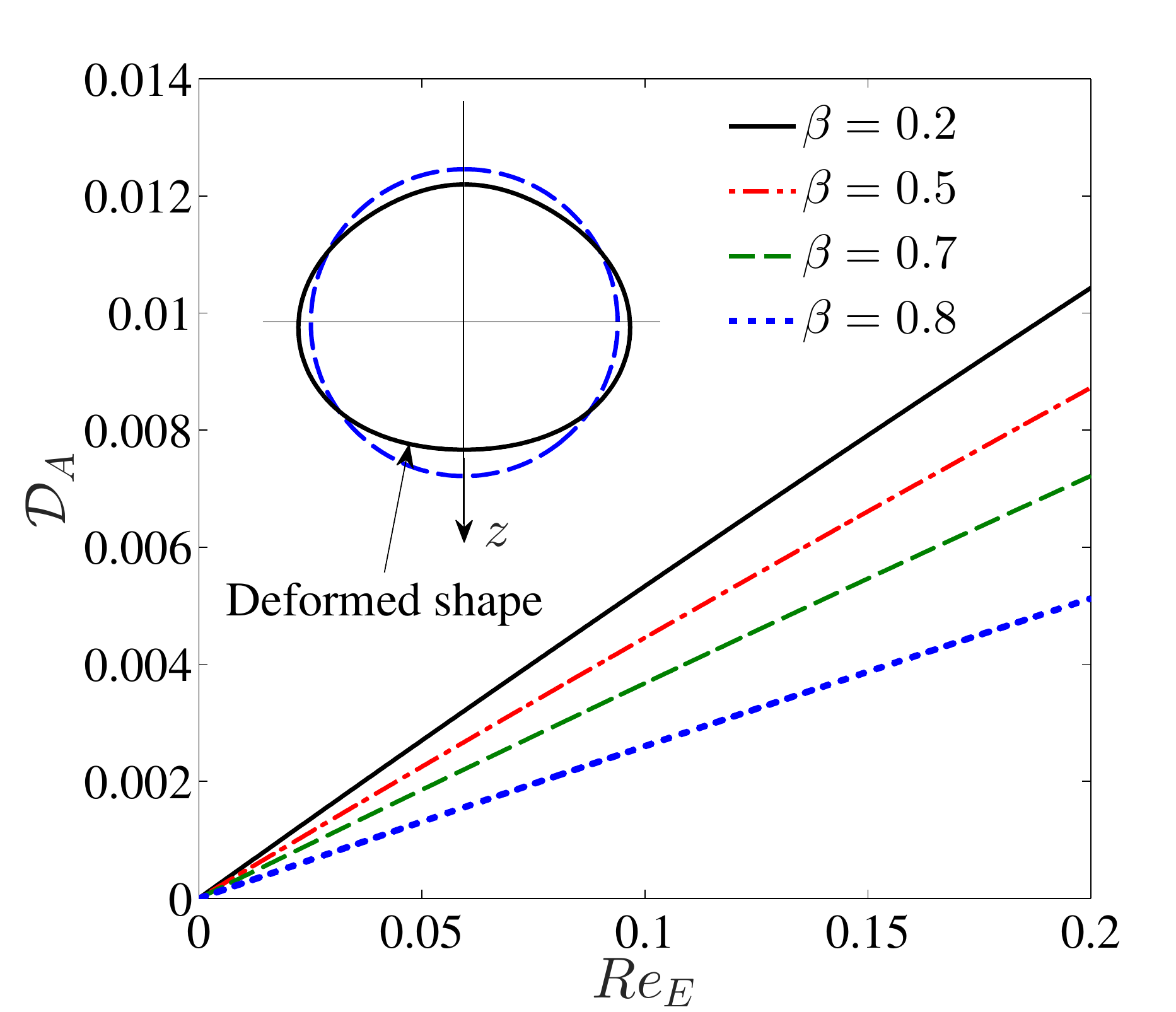}
%				\vspace*{4em}
		\caption{}
		\label{fig:asym_vary_beta}
	\end{subfigure}
	\quad %Horizontal spacing commands.enskip, quad, qquad leave a horizontal space of respectively half an em, one em and two ems. The figure\,\,\ref{fig:D-vs-Ca-vary-beta}"em" is a font depending length, frequently as wide as a capital M in the current font.
	\begin{subfigure}[!htbp]{0.47\textwidth}
		\centering
		\includegraphics[width=\textwidth]{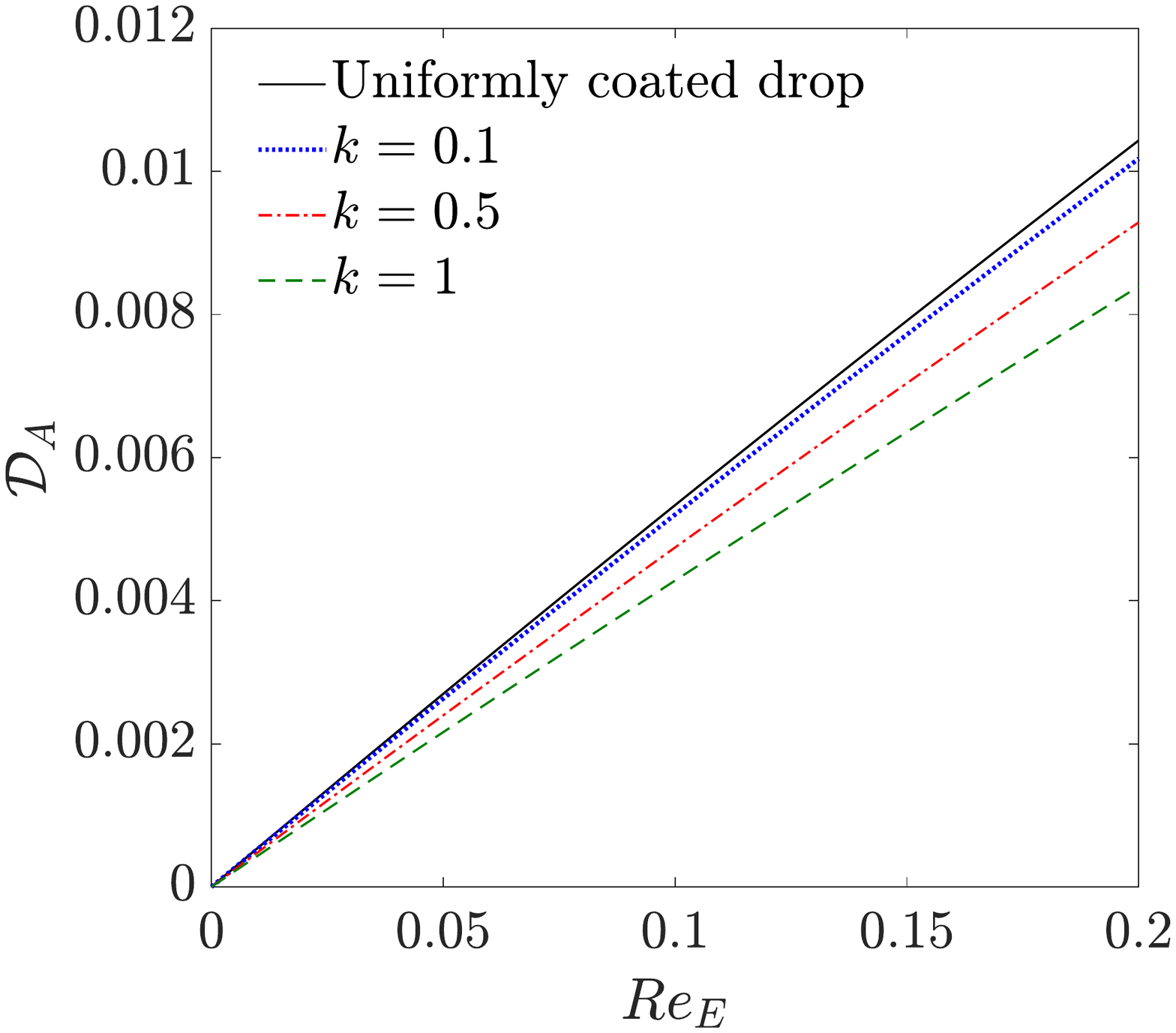}
		\caption{}
		\label{fig:asym_vary_k}
	\end{subfigure}

	\caption{Asymmetric deformation vs. $ Re_{\!_E} $ for (a) various $\beta$ and (b) various $k$.}
	\label{fig:asym-deform-ReE}
\end{figure}

An important aspect of the drop deformation characteristics is that the drop shape becomes asymmetric about the equator. Such an asymmetry is typical for a drop settling with electrohydrodynamic modulation even if the surfactant effects are not present \citep{Spertell1974,Xu2006}. Such behaviour originates from the antisymmetry of the $ L^{(CaRe_{\!_E})}_3P_3(\cos{\theta}) $ term present in  the expression of $ f^{(CaRe_{\!_E})} $ as \eqref{eq:shapefunc_ReE}.
	 To capture such asymmetry  we define a new deformation parameter as \cite{Xu2006}
	\begin{equation} \label{eq:deformation-asym-def}
	\mathcal{D}_A=\dfrac{r_{\!_S}\big|_{\theta=0}-r_{\!_S}\big|_{\theta=\pi}}{r_{\!_S}\big|_{\theta=0}+r_{\!_S}\big|_{\theta=\pi}}.
	\end{equation} 
	In figure\,\,\ref{fig:asym-deform-ReE} we show the effect of surfactant on the asymmetric deformation about the equator triggered by charge convection for a deformable drop.
	The present study reveals that the asymmetric behaviour of shape deformation gets suppressed for increasing values of both the elasticity parameter $ \beta $ (figure\,\,\ref{fig:asym_vary_beta}) and the property ratio $ k $ (figure\,\,\ref{fig:asym_vary_k}).

Based on the electrical conductivity ratio ($ R $) and permittivity ratio ($ S $) some interesting limiting conditions can be deduced in this case. When the $ R\to\infty $ limit is reached the drop behaves as a perfectly conducting one. Such a condition closely resembles various  experimental situations as in \citet{Mhatre2013} and \citet{Ervik2018} where aqueous drops are considered in an oil medium. In such conditions the electric field acts perpendicularly to the drop surface and the tangential electric field components becomes identically zero, having no effect on $ O(Re_{\!_E}) $  velocity perturbation. Hence such a limit can alter only the $ O(Ca) $ correction to drop velocity which is obtained as

\begin{equation}
\begin{split}
&\textrm{Low P\'eclet number limit:}\\
&\lim_{R\to\infty} U^{(Ca)}_d={\frac { \left( 9\,{\lambda}^{2}-3\,\lambda+24 \right) M \left( 
		2+3\,\lambda \right) }{ 20\left( -3\,\lambda-2 \right) ^{2} \left( 1+
		\lambda \right) }}
\\
&\qquad \qquad \quad \, \, \, \, \, -\underbrace{{\frac { \left(  \left( 108\,{\lambda}^{2}-15\,\lambda\,k
			-36\,\lambda+20\,k-72 \right) \beta-108\,{\lambda}^{2}+36\,
			\lambda+72 \right) \beta\,kM}{20 \left( 2+3\,\lambda \right) 
			\left(  \left( 3\,\lambda-k+2 \right) \beta-3\,\lambda-2
			\right) ^{2} \left( 1+\lambda \right) }}}_\text{correction due to surfactant}
\\
&\textrm{High P\'eclet number limit:}\quad \\
&\lim_{R\to\infty} U^{(Ca)}_d={\frac { \left( 9\,{\lambda}^{2}-3\,\lambda+24 \right) M \left( 
		2+3\,\lambda \right) }{ 20\left( -3\,\lambda-2 \right) ^{2} \left( 1+
		\lambda \right) }}
+\underbrace{{\frac {\left( 3\,\lambda-4 \right) M}{4 \left( 1+\lambda
			\right)  \left( 2+3\,\lambda \right) }}}_\text{correction due to surfactant}
\end{split}
\label{eq:Ca-order-Ud_PC}
\end{equation}

\subsection{Perfectly dielectric media}
\label{subsec:pdm}

\begin{figure}[!htbp]
	\centering
	\begin{subfigure}[!htbp]{0.45\textwidth}
		\centering
		\includegraphics[width=\textwidth]{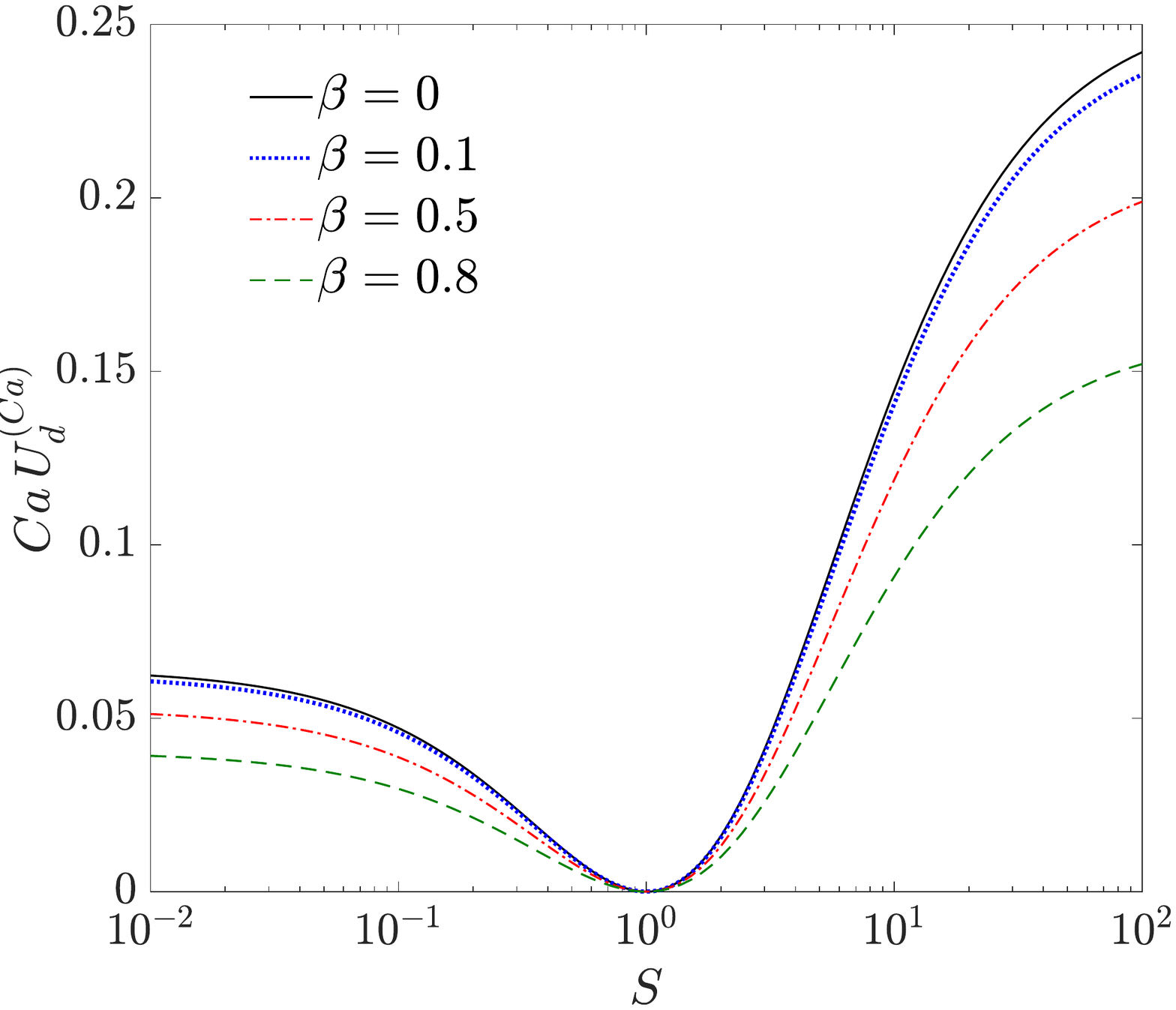}
		\caption{}
		\label{fig:LD-corr-vary-beta}
	\end{subfigure}
	\quad %Horizontal spacing commands.enskip, quad, qquad leave a horizontal space of respectively half an em, one em and two ems. The "em" is a font depending length, frequently as wide as a capital M in the current font.
	\begin{subfigure}[!htbp]{0.45\textwidth}
		\centering
		\includegraphics[width=\textwidth]{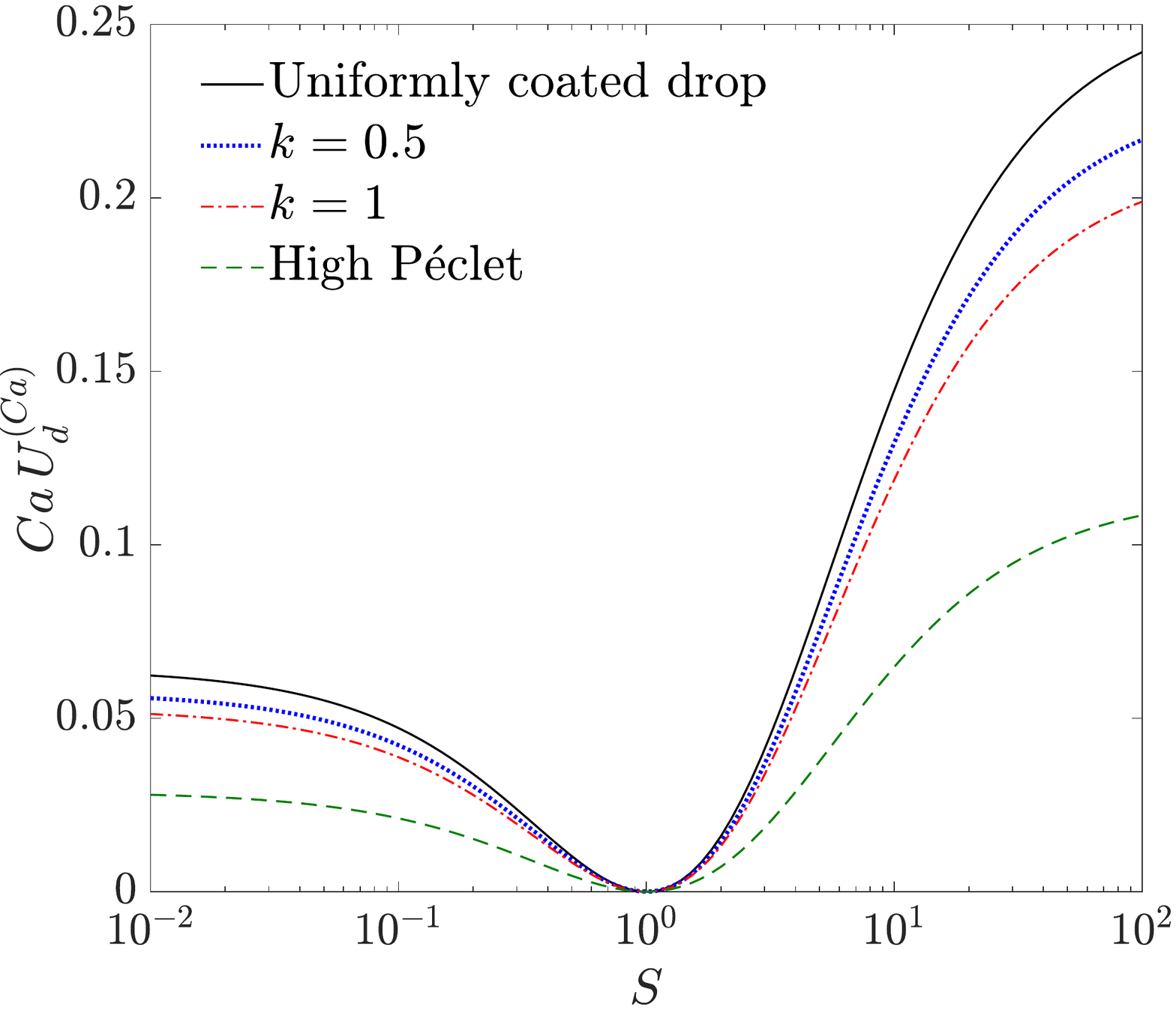}
		\caption{}
		\label{fig:LD-corr-vary-k}
	\end{subfigure}

	\caption{$O(Ca)$ velocity correction vs. permittivity ratio for (a) various $\beta$ with $ k=1 $ and (b) various $k$ with $ \beta=0.5 $. The other parameters are $ M=5, Ca=0.2 $ and $\lambda=0.4359$.}
	\label{fig:LD-corr-vary-k-beta}
\end{figure}

Another important case of ($ R,S $) combination often employed in EHD studies is when a perfectly dielectric drop is settling in another perfectly dielectric medium. This limit can be mathematically deduced by taking $ R \to S $ limit. Similar to the perfectly conducting drop case here  also the $ U^{(Re_{\!_E})}_d $ vanishes due to zero surface charge in the leading-order as well as in $ O(Re_{\!_E}) $. However the scenario gets changed when the drop deforms and the expression for $ O(Ca)$ drop velocity for such situations can be obtained as below:
\begin{equation}
\begin{split}
&\textrm{Low P\'eclet limit:}\\
&\lim_{R \to S} U^{(Ca)}_d={\frac {3\,M \left( 3\,{\lambda}^{2}-\lambda+8 \right)  \left( S-1
		\right) ^{2}}{ \left( 20+20\,\lambda \right)  \left( 2+3\,\lambda
		\right)  \left( 2+S \right) ^{2}}}
\\
& \, \, \, \, \, -\underbrace{{\frac { \left( S-1 \right) ^{2} \left( 108\,{\lambda}^{2}\beta
			-15\,k\lambda\,\beta-36\,\lambda\,\beta+20\,k\beta-108\,{
				\lambda}^{2}-72\,\beta+36\,\lambda+72 \right) \beta\,kM}{20
			\left( 2+3\,\lambda \right)  \left( 2+S \right) ^{2} \left( 3\,
			\lambda\,\beta-k\beta+2\,\beta-3\,\lambda-2 \right) ^{2} \left( 
			1+\lambda \right) }}}_\text{correction due to surfactant}
\\
&\textrm{High P\'eclet limit:}\quad \\
&\lim_{R \to S} U^{(Ca)}_d={\frac {3\,M \left( 3\,{\lambda}^{2}-\lambda+8 \right)  \left( S-1
		\right) ^{2}}{ \left( 20+20\,\lambda \right)  \left( 2+3\,\lambda
		\right)  \left( 2+S \right) ^{2}}}
+\underbrace{{\frac { \left( S-1 \right) ^{2} \left( 3\,\lambda-4 \right) M}{4
			\left( 1+\lambda \right)  \left( 2+3\,\lambda \right)  \left( 2+S
			\right) ^{2}}}}_\text{correction due to surfactant}
\end{split}
\label{eq:Ca-order-Ud_LD}
\end{equation}
In figure\,\,\ref{fig:LD-corr-vary-k-beta} the effect of variation of the permittivity ratio on the drop velocity has been highlighted for various $ \beta $ and $ k $. It shows that the retardation effect of surfactant on the $ O(Ca) $ correction to drop velocity increases if either the parameter $ S $ is increased or decreased from the point $ S=1 $. The vanishing value of this perturbation velocity at $ S=1 $, which is evident from the \eqref{eq:Ca-order-Ud_LD}, is also portrayed in both the figures \ref{fig:LD-corr-vary-beta} and \ref{fig:LD-corr-vary-k}. 
Since in this specific case,  electrical properties of both the fluids become the same ($ R=S=1 $), the electric Maxwell stress jump at the interface vanishes. As a consequence charge accumulation on the surface does not take place resulting in zero shape deformation. Thus the settling velocity reduces to the classical Hadamard - Rybczynski velocity only with a corresponding correction due to surfactant ($U^{(0)}_d$).  

\section{Conclusions}

In the present work, we have investigated the alterations caused due to the  presence of surface-active agents on the sedimentaion of a drop which is simultaneously driven by gravity and an uniform electric field in the direction of gravity. Assuming a low concentration of surfactant molecules on the fluid-fluid interface and a bulk insoluble nature, we have solved the  surfactant transport equation for two plausible physical limits (low and high $ Pe $) which are determined by considering the relative importance of the governing mechanisms involved. The  electric field effect on the hydrodynamics has been modeled as per the leaky dielectric model. To consider situations of practical relevance, we have taken into account both the surface  charge convection phenomena and deformable nature of the drop shape. Subsequently, we obtained analytical solutions of the electric potential and velocity field using a double asymptotic perturbation in terms of small capillary number ($ Ca $) and a small electric Reynolds number ($ Re_{\!_E} $). For two different limiting conditions, the present results show an excellent agreement with the existing theoretical and numerical works. Although the results have been mainly demonstrated for a leaky dielectric drop in another leaky dielectric medium,  corresponding behaviours in other limits with respect to the electrical parameters ($ R,\,S $) have also been  highlighted in some of the specific cases. The major conclusions  that can be drawn from the present study are summarized below:

\begin{enumerate}[label=(\roman*),leftmargin=0pt,itemindent=3em]
	\item 
	The results indicate that the surfactant distribution can modify the charge convection phenomena through the opposing Marangoni flow in such a way that the location of stagnation points gets shifted. This alters the internal vortex structures severely. For specific electrical parameter combinations, the Marangoni flow has the ability to completely suppress the additional  secondary roles which otherwise come into existence due to charge convection in case of a clean drop.  In the high P\'eclet limit, the droplet surface becomes immobile in nature and the charge convection effect vanishes altogether. Such a behaviour can serve as a means for  better control of mixing characteristics. 
	\item 
For the case of spherical drop,  the drop settling velocity $ U_{d} $ gets reduced with an increase in elasticity parameter $ \beta $ in the low $ Pe $ limit. It is also found that the correction to the drop settling velocity  due to surface charge convection   ($ U^{(Re_{\!_E})}_{d} $) is always reduced for any combination of surfactant parameters $ \beta, \,k $, in the presence of surfactant. An increase in both the parameters $ k $ and $ \beta $ causes an increase the surface tension gradient which in turn enhances the drag force opposing the drop motion. Thus, similar to $ \beta $, $ k $  causes a dip in the drop settling  velocity. However, in the high $ Pe  $ limit, the electric effects can have no role to play in $ U_d $ due to a vanishingly small surface velocity. 
	\item 
	When the drop shape deforms, the Marangoni inhibition on the flow field is found to be dominated by the tip-stretching effect. As a consequence, the deformation of drop shape gets enhanced with increased sensitivity of surface tension to surfactant concentration (increasing $  \beta $) as well as with rising values of  the property  ratio ($ k $). For the system-A presented, the drop becomes more and more oblate in shape, thus intensifying the flow obstruction. This causes a corresponding reduction in the drop velocity. However for the prolate drop shape (system-B), similar competing mechanisms of surfactant transport causes further escalation in drop settling velocity.
	\item
	 The asymmetric pattern of the deformed drop shape is captured in the present analysis. Both the surfactant parameters reduce the  asymmetric deformation parameter ($ \mathcal{D}_A $). For higher values of $ Re_{\!_E} $, the surfactant effect on $ \mathcal{D}_A $ becomes increasingly prominent.
	\item 
	 In case of perfectly dielectric drop falling in another perfectly dielectric medium, the velocity correction  due to deformation vanishes for equal permittivity of the fluid pair. In the situations where either $ S<1 $ or $ S>1 $ condition is met, Marangoni stress increases the drop velocity.
\end{enumerate}

\begin{appendices}
	\renewcommand{\thesection}{\Alph{section}}
	\renewcommand{\thesubsection}{\thesection.\arabic{subsection}}
	\renewcommand\thefigure{A-\arabic{figure}}    
	\setcounter{figure}{0} 
	\renewcommand{\theequation}{A \arabic{equation}}
	% redefine the command that creates the equation no.
	\setcounter{equation}{0}  % reset counter 
\section{}
\label{App-validation}

In this section the following two limiting cases will be considered to check the validity of the present analytical results:
\subsection*{(i) Settling of a clean drop in presence of electric field: }  \label{Apss:settling-no-surf}
In this case the driving forces for drop motion are both the gravity and electrohydrodynamic force. The drop settling velocity and deformation in the presence of charge convection, for such a system,  were obtained analytically by \cite{Xu2006}. They also obtained agreement of their solutions with experimental observations. However they considered clean drops with uniform surface tension along the interface. To realize such a limiting condition, if a  substitution $\beta=0$ is made, the resulting expressions of different harmonics of electric potential, streamfunction, shape function and drop settling velocity reduce to  the exact ones as obtained by \citet{Xu2006}. 
\subsection*{(ii) Neutrally buoyant drop with surfactant effects: }\label{Apss:neutrally-buoyant-surf}
%\cite{Nganguia2013,Teigen2010,Ha1998}:
In such a condition the drop is treated as a stationary suspended one ($ \rho_i=\rho_e $) with no effect of the buoyancy force \citep{Nganguia2013,Teigen2010,Ha1998,Ha1995}. Moreover charge convection was not considered to have significant effect.
Such a limit can be retrieved from the present study, if we let $M=1, \mathbf{F}_B=0$ and take the electrohydrodynamic velocity scale ($\epsilon_eaE^2/\mu_e$) as a base for non-dimensionalization.

In the the above limiting situation, a deeper reflection on the solution of leading-order surfactant transport equation shows that the the $ \Gamma^{(Ca)}_1 $ harmonic vanishes in \eqref{eq:surf-Ca} and the expression of $ \Gamma^{(Ca)}_2 $

\begin{equation}\label{eq:match-HY}
\Gamma^{(Ca)}_2={\frac {3\,k_{{1}} \left( -1+\beta \right)  \left( R-S \right) }{
		\left( 5\,\beta\,\lambda-\beta\,k_{{1}}+5\,\beta-5\,\lambda-5
		\right)  \left( R+2 \right) ^{2}}},
\end{equation}
which matches with the theoretical investigation performed by \cite{Ha1995} in the diffusion dominated limit ($ k\sim O(1) $). It is to be noted here that although identical symbols are used for the elasticity parameter ($ \beta $) and the conductivity ratio $ (R) $ for both cases, the definitions are different as $ \beta={\dfrac {\beta_{{ HY}}}{1+\beta_{{HY}}}} $ and $ R=\dfrac{1}{R_{HY}} $. Similar to the surfactant concentration the $ O(Ca) $ degree of deformation ($ \mathcal{D} $) also reduces to the expression obtained by them.
An investigation of different harmonics in shape deformation, it is found that the effect of buoyancy is conveyed only through the harmonic $L^{(Ca^2)}_3$ present in the $ O(Ca) $ solution. In the present study, under the limiting simplifications as discussed above, the physical situation resembles \citet{Nganguia2013} in the non-diffusing regime ($Pe\to\infty$). 
In order to validate our calculations under, this simplified case, we present  the deformation parameter ($\mathcal{D}$)  obtained from the present study
\begin{equation}\label{key}
\mathcal{D}=\frac {9\left( {R}^{2}+2R-4S+1 \right) }{16
	\left( R+2 \right) ^{2}}\,Ca+{\frac {  139{R}^{3}+264{R}^{2}-696RS+111R+336S-154 }
	{80 \left( R+2 \right) ^{3}}}\,{Ca}^{2},
\end{equation}
which matches exactly with the small-deformation theory  result of  \citet{Nganguia2013}. 
%\textcolor{red}

\begin{figure}[!htbp]
	\centering
	\begin{subfigure}[!htbp]{0.45\textwidth}
		\centering
		\includegraphics[width=\textwidth]{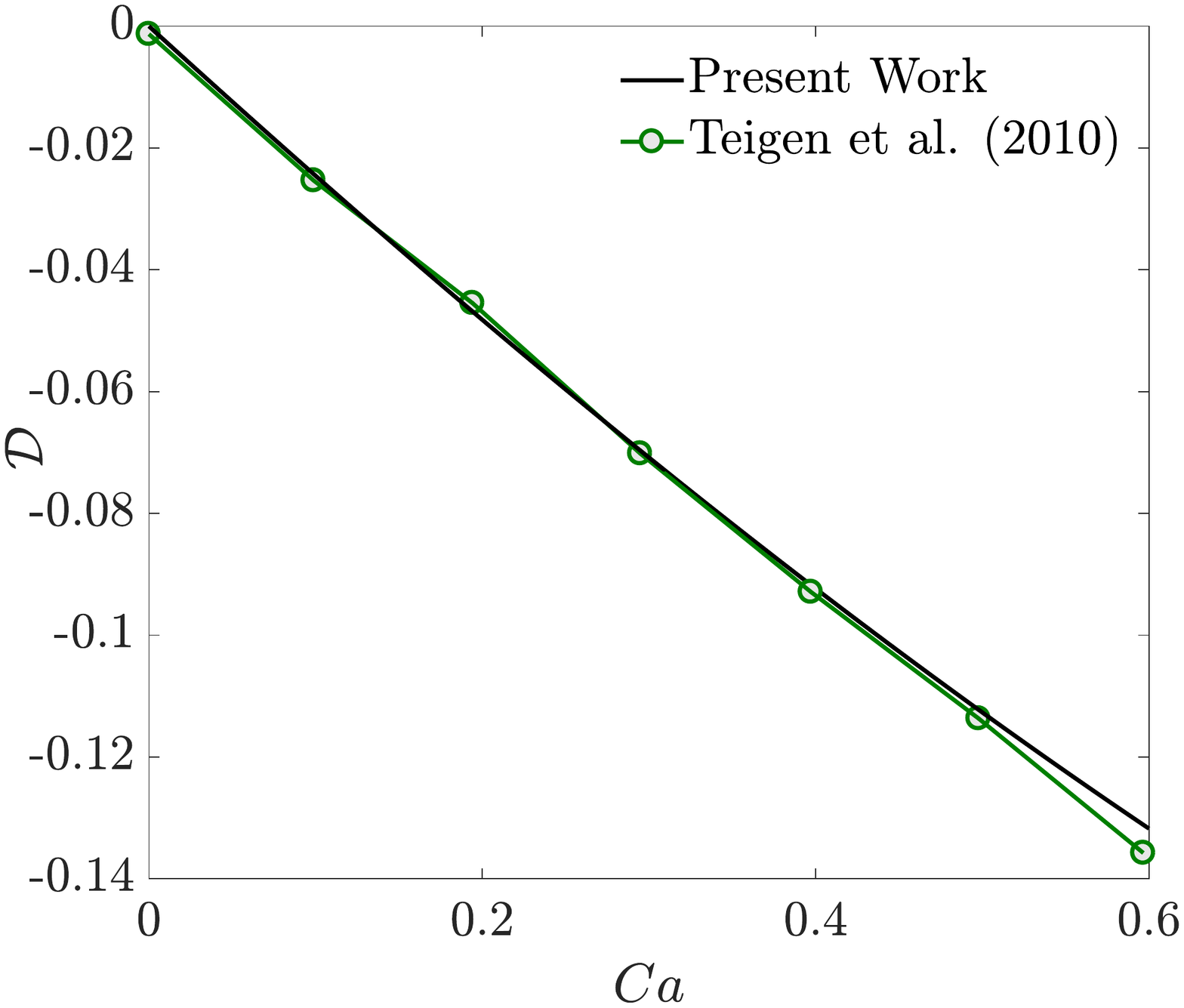}
		\caption{}
		\label{fig:Teigen_validation_4c}
	\end{subfigure}
	\quad %Horizontal spacing commands.enskip, quad, qquad leave a horizontal space of respectively half an em, one em and two ems. The "em" is a font depending length, frequently as wide as a capital M in the current font.
	\begin{subfigure}[!htbp]{0.45\textwidth}
		\centering
		\includegraphics[width=\textwidth]{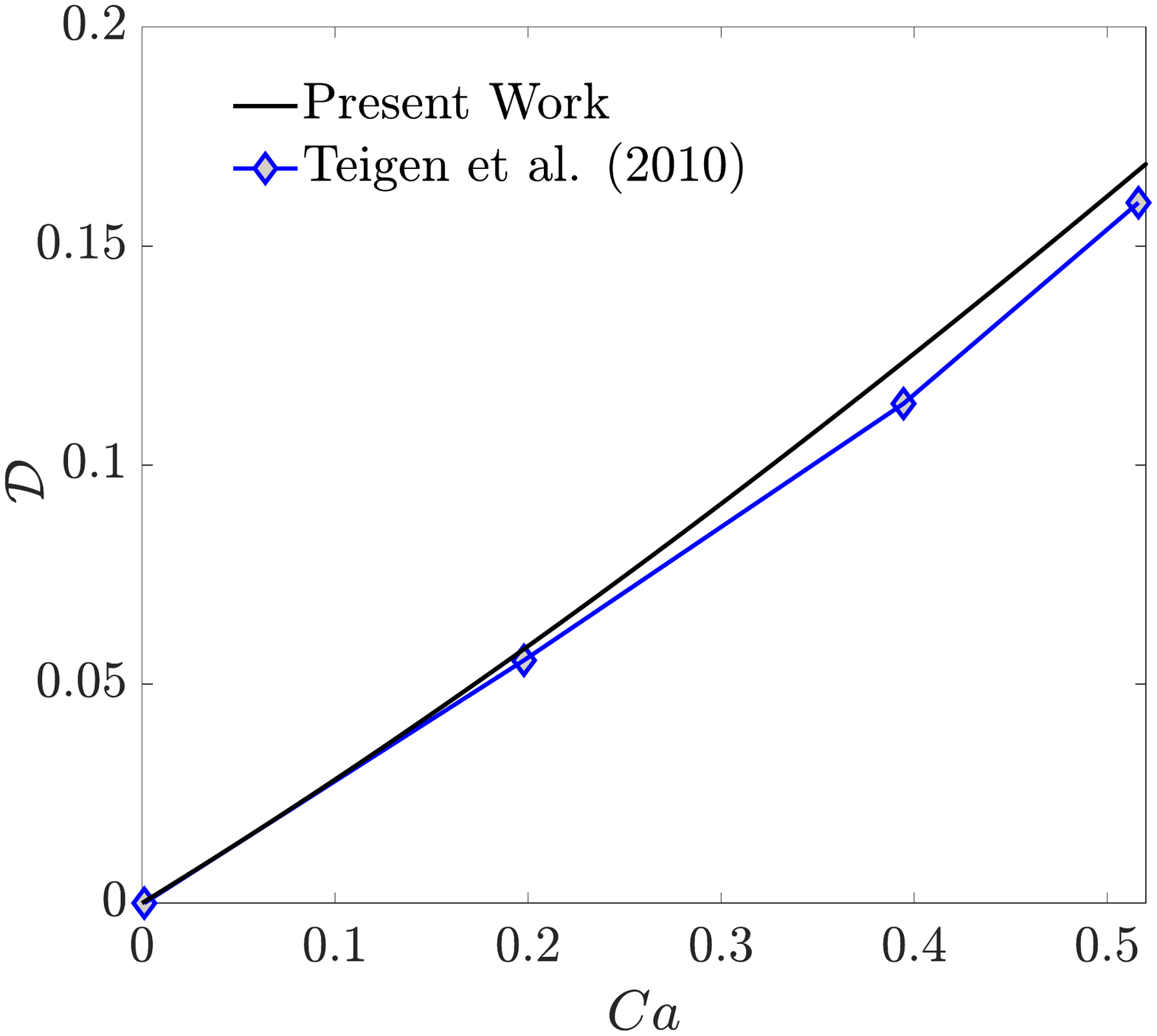}
		\caption{}
		\label{fig:Teigen_validation_4a}
	\end{subfigure}
	\caption{Comparison of deformation variation with the previous numerical investigation by \protect{\citet{Teigen2010}} for a limiting case of a neutrally buoyant droplet. The choice of physical parameters pertinent to individual cases are: (a) $R=1,\,S=2,\, \lambda=1$ and (b)  $R=3,\, S=1,\, \lambda=1$.}
	\label{fig:Teigen-validation}
\end{figure}

In their work \citet{Teigen2010} has performed numerical simulations using a level-set method, for a similar condition with a surface P\'eclet number of 10 ( $ Pe=10 $). In that limit, our calculations reach the high P\'eclet limit ($k\sim O(100)$).The comparisons of the deformation parameter with their work for two different combinations of electrical properties are shown in Figure \ref{fig:Teigen-validation}. It is observed that both for oblate (figure\,\,\ref{fig:Teigen_validation_4c}) and prolate (figure\,\,\ref{fig:Teigen_validation_4a}) kind of deformation, the present small deformation theory matches well with their numerical predictions.

\section{}
\label{sec:App-1}
\begin{figure}[!htbp]
	\centering
	\begin{subfigure}[!htbp]{0.48\textwidth}
		\centering
		\includegraphics[width=\textwidth]{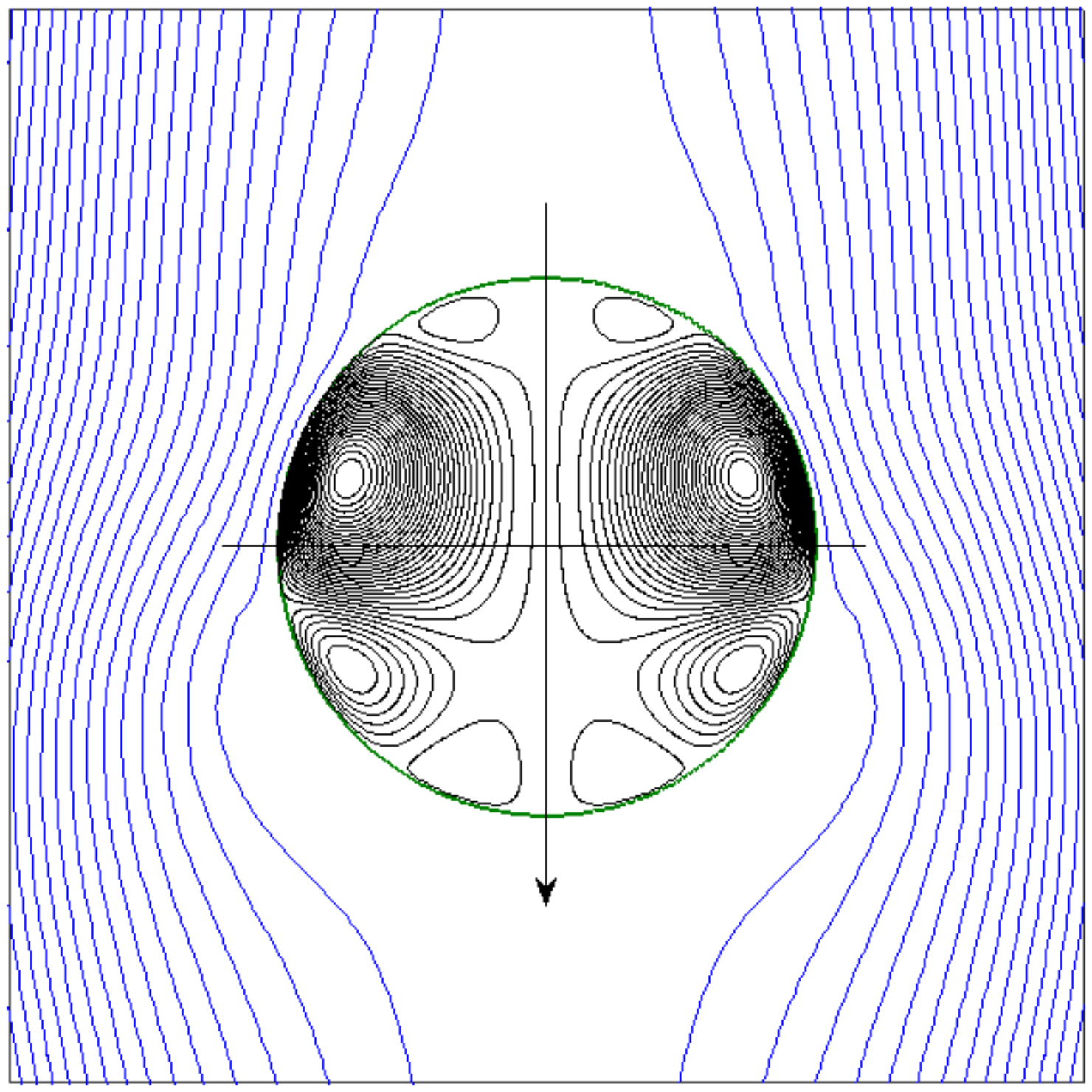}
		\caption{Clean drop}
		\label{fig:Xu_contour}
	\end{subfigure}
	%	\quad %Horizontal spacing commands.enskip, quad, qquad leave a horizontal space of respectively half an em, one em and two ems. The "em" is a font depending length, frequently as wide as a capital M in the current font.
	\hfill
	\begin{subfigure}[!htbp]{0.48\textwidth}
		\centering
		\includegraphics[width=\linewidth]{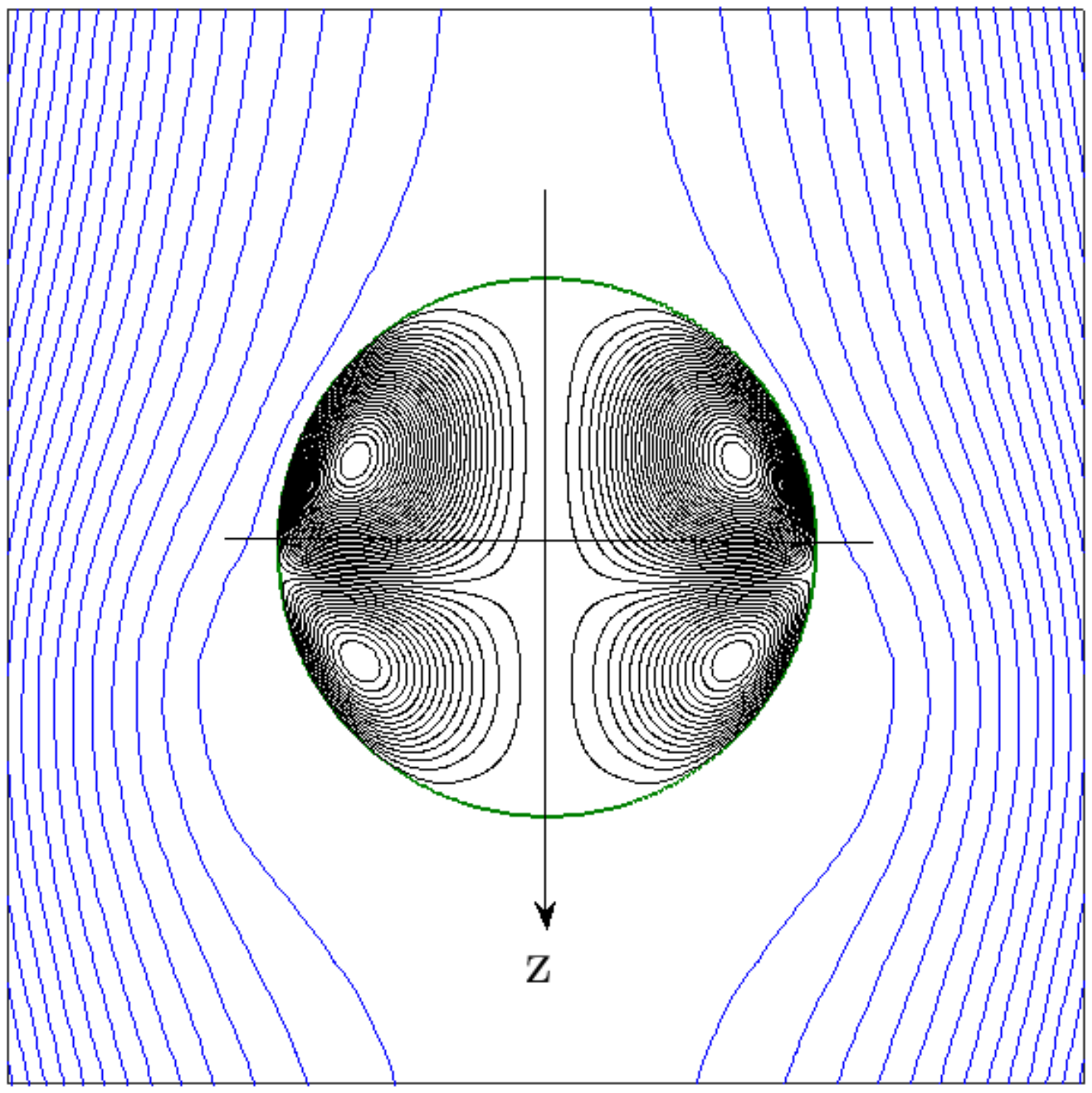}
		\caption{Surfactant-laden drop ($ \beta=0.8,\, k=1 $)}
		\label{fig:Xu_contour_surf}
	\end{subfigure}
	\\[-6ex] %Add something like this just after the second subfigure reduces the distance between the first and second rows.
	\begin{subfigure}[!htbp]{0.48\textwidth}
		\centering
		\includegraphics[width=\linewidth]{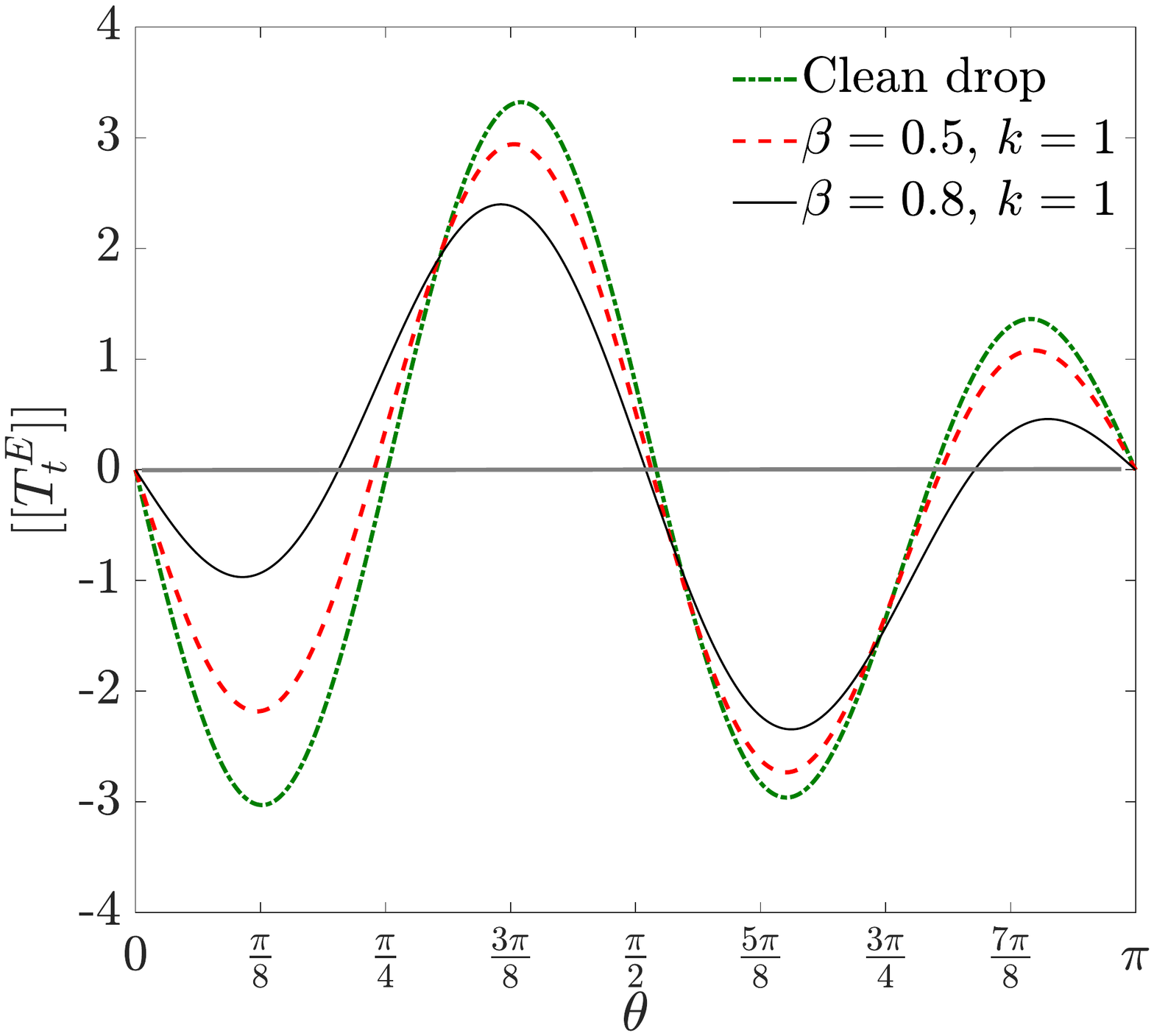}
		\caption{Variation of $ \llbracket T^E_t\rrbracket $ with $\beta $}
		\label{fig:Homsy_TE_t_vary_beta}		
	\end{subfigure}
	\begin{subfigure}[!htbp]{0.48\textwidth}
		\centering
		\includegraphics[width=\linewidth]{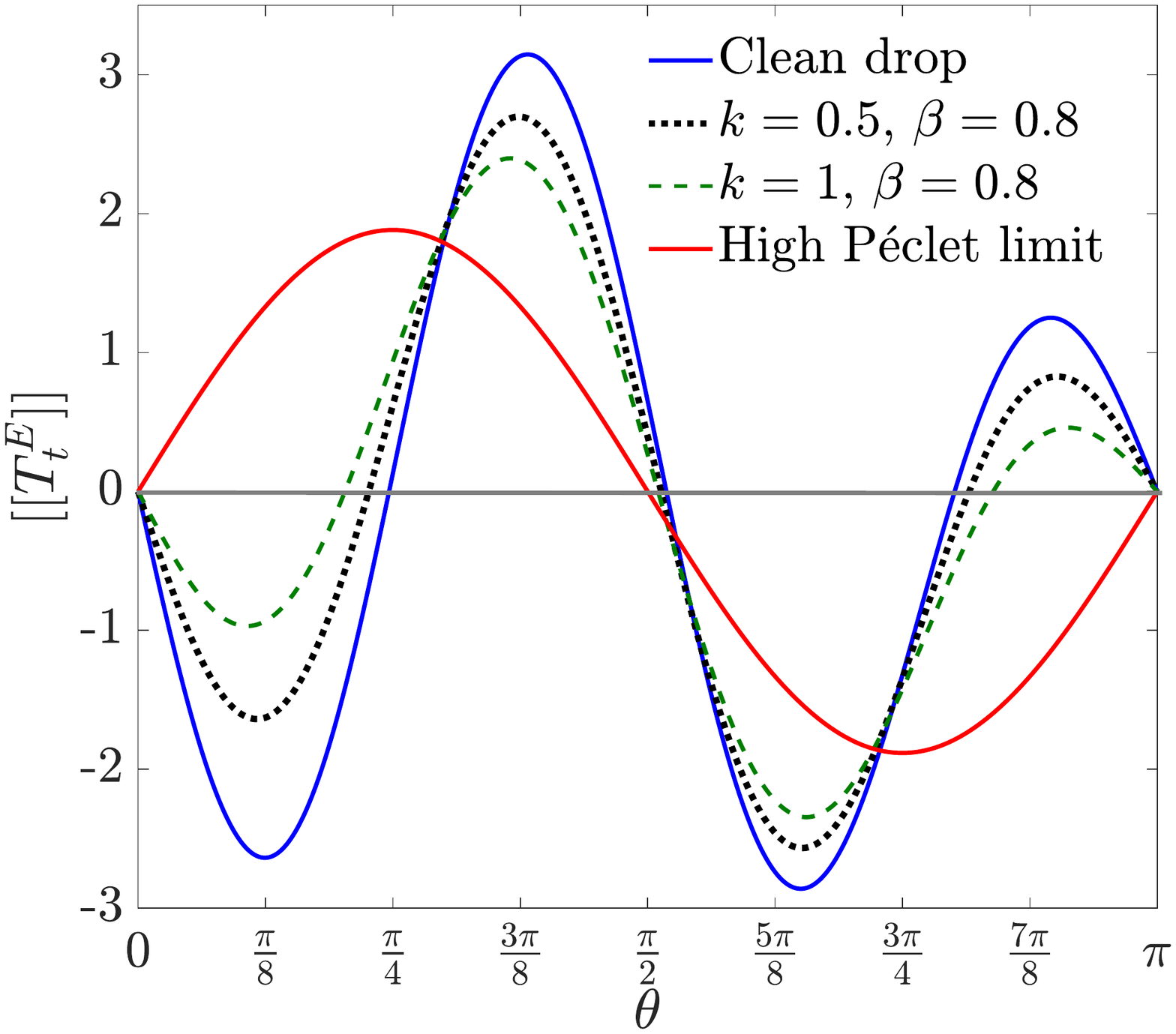}
		\caption{Variation of $ \llbracket T^E_t\rrbracket $ with $k $}
		\label{fig:Homsy_TE_t_vary_k}			
	\end{subfigure}
	\vspace*{30pt}
	\caption{Streamline pattern for (a) clean drop with property values used in \protect{\citep{Xu2006}} and (b) for surfactant coated case of the same drop. Subplot (c) is the tangential electric stress jump for different values of $ \beta $ and (d) is for different values of  $ k $.}
	\label{fig:Xu-Homsy-Streamilne-Stress}
\end{figure}

In this section the streamline pattern and electrical tangential stress jump for $ R=0.01,S=1.7  $, $ \lambda=0.03  $ and $ M=2.7 $ are presented in Fig~\ref{fig:Xu-Homsy-Streamilne-Stress}.

\section{ }\label{sec:oblate-shape-deform}
\begin{figure}[!htbp]
	\centering
%			\vspace*{-12em}
	\begin{subfigure}[!htbp]{0.52\textwidth}
		\centering
		\includegraphics[width=\textwidth]{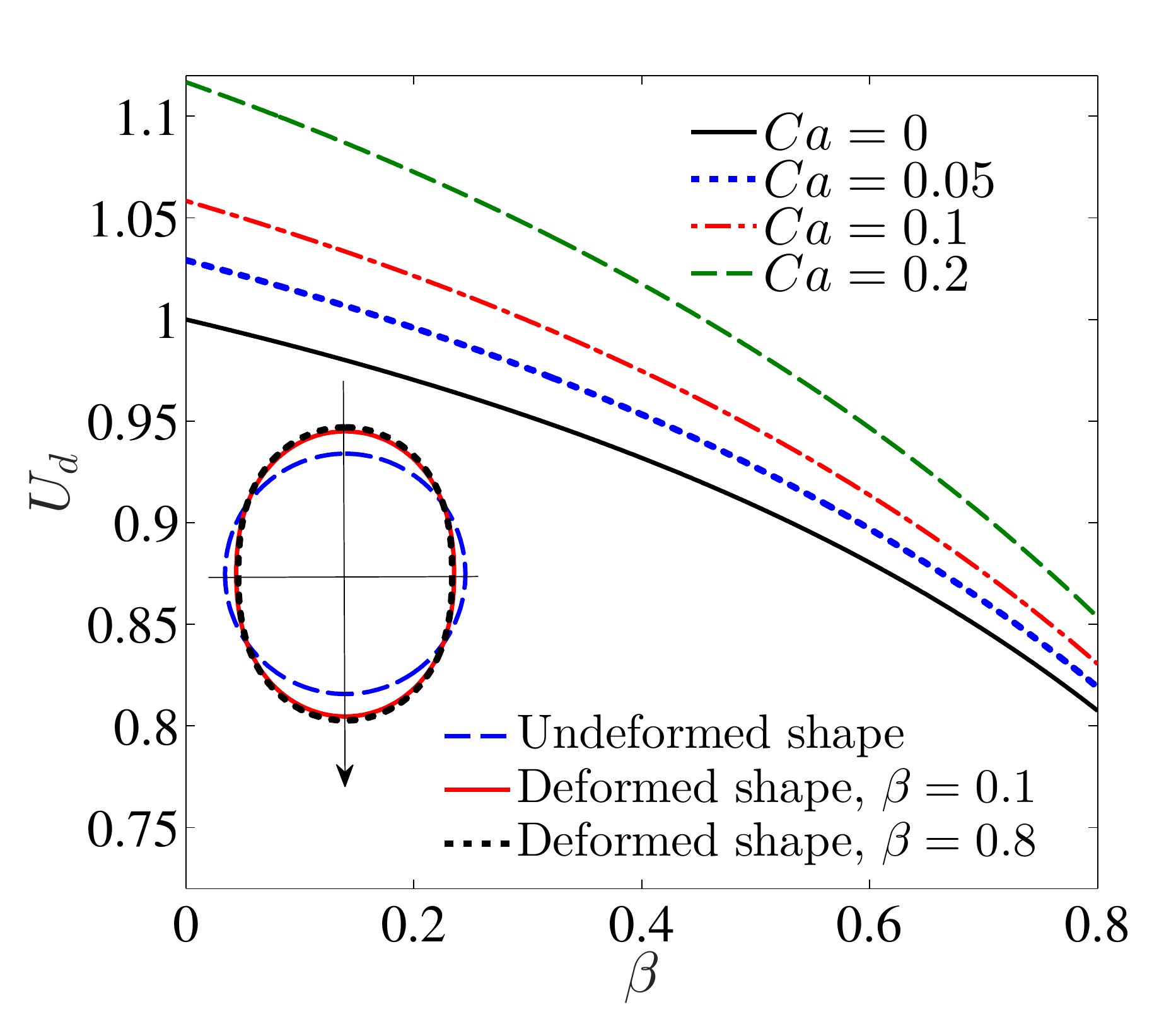}
				\vspace*{5em}
		\caption{}
		\label{fig:prolate-vel-beta-Ca}
	\end{subfigure}
%		\vspace*{-1em}
	\begin{subfigure}[!htbp]{0.47\textwidth}
				\vspace*{-3em}
		\centering
		\includegraphics[width=\linewidth]{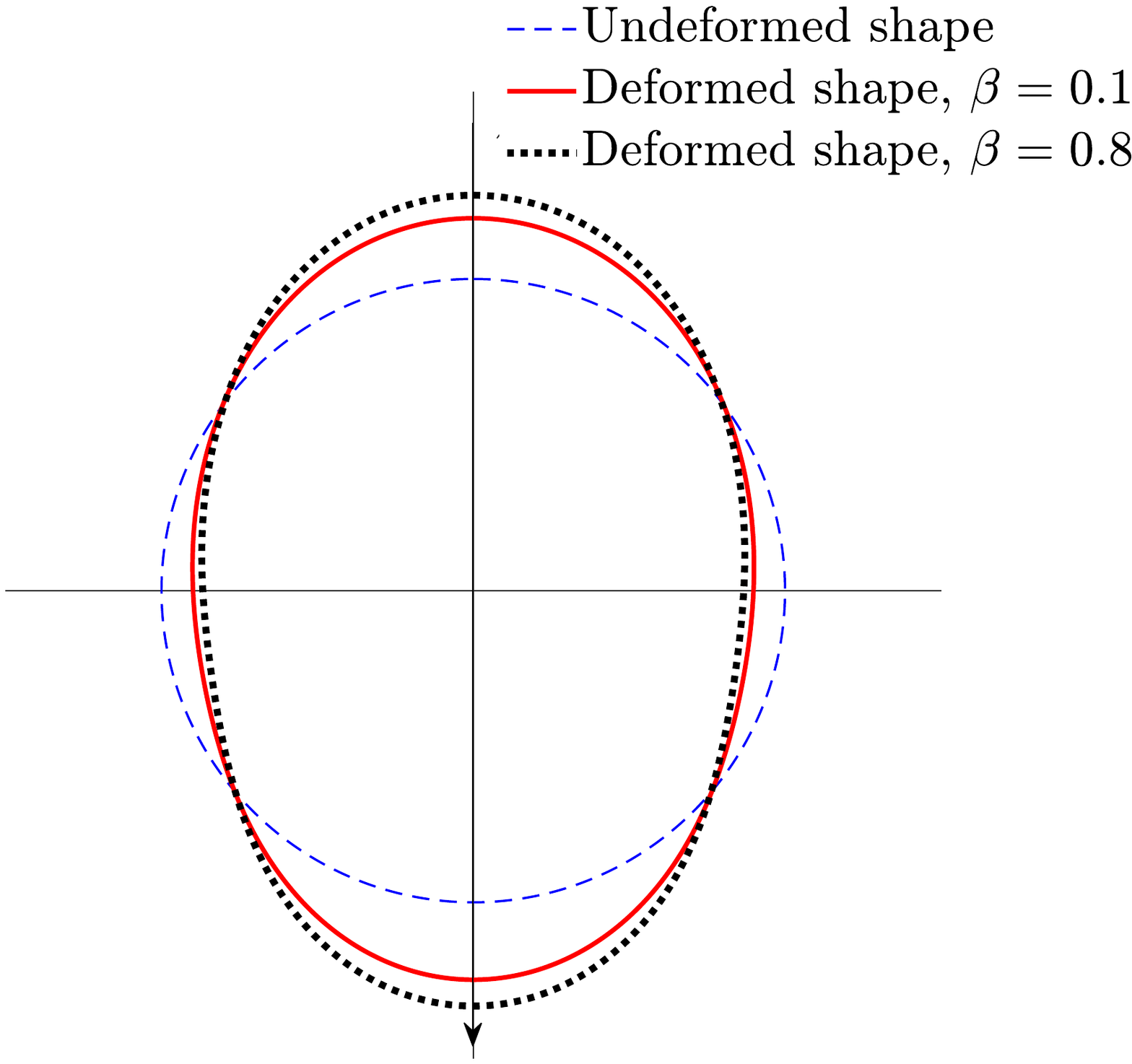}
		\caption{}
		\label{fig:prolate_shape_vary_beta}
	\end{subfigure}
	\caption{Surfactant parameters affecting drop velocity and shape deformation for system-B. Subplot (a) $ U_d$ vs. $\beta$ for different $ Ca $. The chosen parameters are  $ M=2.5$ and $ k=1 $. In the inset the leading-order deformed shapes are shown for various $\beta  $ and $ Ca=0.2 $. Subplot (b) shows shapes with higher order deformation and different $ \beta $ .}
	\label{fig:prolate-deform}
\end{figure}

Here we show the surfactant effects on drop velocity and shape deformation characteristics for system-B (prolate drop) in figure~\ref{fig:prolate-deform}.
\end{appendices}

%\FloatBarrier  %to force bibliography at the end
%\bibliography{EHD_surf_only_this.bib}

\end{document}